\documentclass[authoryear,12pt]{elsarticle}      
\usepackage[english]{babel}
\usepackage{multirow}
\usepackage{amsmath}
\usepackage[toc,page]{appendix}
\usepackage{amssymb}        
\usepackage{amsthm}         
\usepackage{natbib}         
\usepackage{graphicx}       
\usepackage{subfigure}      
\usepackage{color}
\journal{Journal of the Mechanics and Physics of Solids. Accepted for publication.}

\begin{document}
\begin{frontmatter}
\title{High fidelity simulation of the mechanical behavior of closed-cell polyurethane foams}
\author{M. Marvi-Mashhadi$^{1,2}$}
\author{C. S. Lopes$^{1}$}
\author{J. LLorca$^{1,2}$}

\address{$^1$ IMDEA Materials Institute, C/ Eric Kandel 2, 28906, Getafe, Madrid, Spain. \\  $^2$ Department of Materials Science, Polytechnic University of Madrid/Universidad Polit\'ecnica de Madrid, E. T. S. de Ingenieros de Caminos. 28040 - Madrid, Spain.}

\begin{abstract}

The mechanical behavior of closed-cell foams in compression is analyzed by means of the finite element simulation of a representative volume element of the microstructure. The digital model of the foam includes  the most relevant details  of the microstructure (relative density, cell size distribution and shape, fraction of mass in the struts and cell walls and strut shape), while the numerical simulation takes into account the influence of the gas pressure in the cells and of the contact between cell walls and struts during crushing. The model was validated by comparison with experimental results on isotropic and anisotropic polyurethane foams and it was able to reproduce accurately the initial stiffness, the plateau stress and the hardening region until full densification in isotropic and anisotropic foams. Moreover, it also provided good estimations of the energy dissipated and of the elastic energy stored in the foam as  a function of the applied strain.   Based on the simulation results, a simple analytical model was proposed to predict the mechanical behavior of closed-cell foams taking into the effect of the microstructure and of the gas pressure.  An example of application of the simulation tool is presented to design foams with an optimum microstructure from the viewpoint of energy absorption for packaging.

\end{abstract}


\end{frontmatter}

\section{Introduction}

The mechanical properties of foams can be tailored to meet a wide range of mechanical  (stiffness, flow stress, energy absorption and recovery, etc) and functional properties (thermal and acoustic insulation) by carefully tailoring the microstructural features (density, cell size and shape, fraction of solid material in cell walls and struts, etc.) and the properties of the solid material \citep{Berlin1980, GibsonBookCellular}. As a result, foams are widely used in many industrial sectors  and, among them, polyurethane (PU) foams stand out for their wide use in impact-friendly surfaces, energy absorbing structural components, packaging material, etc. \citep{Mills2007xvii, Avalle2001455, Daniel2009}.

The versatility of foams to achieve a wide range of properties (stiffness, strength, energy absorption, etc) with the same bulk material is determined by the complex microstructure that allows to tailor the macroscopic response up to a much larger extent than in other solid materials. The link between the microstructure of the foams and the mechanical properties was pioneered by Gibson and Ashby \citep{GibsonBookCellular} who developed analytical models based on the analysis of the behavior of simple unit cells. Based on these results and on extensive experimental data, many phenomenological models have been developed to take into account the influence of strain rate, temperature and density on the mechanical properties of foams subjected to uniaxial compression (see, for instance, \citep{LSG05, ABI07, OA09}). Nevertheless, these models cannot provide information to design foams with optimum properties and do not consider the behavior during unloading-reloading. This latter information is critical to establish the cushioning diagrams that are normally used to design  foams for packaging.

Over the years, more accurate models have been developed to establish the microstructure-properties link in foams. They began by assuming that the foam microstructure was made up by an ensemble of Kelvin cells \citep{Jang20102872,Sullivan20081754,Subramanian25052012}. More recently, accurate descriptions of the microstructure of the foams were created by means of 3D  Voronoi or Laguerre tessellations \citep{TEKOGLU2011109,Redenbach201270, KH16, MLL17} or soap froth models \citep{VRS16}. These 3D digital microstructures  were able to include all the  details of the foam microstructure, such as the strut shape, the average number of faces per cell, the average number of edges per face, etc. The effective properties were obtained through numerical simulation of these representative volume elements (RVEs) of the microstructure \citep{TEKOGLU2011109, Chen2015150, Redenbach201270, KH16, MLL17, MLL17anisotropy}. This strategy was able to account for  the influence of cell size distribution, cell anisotropy, mass distribution between faces and struts, etc.) on the elastic constants on the foam \citep{TEKOGLU2011109, Chen2015150, KH16} and, more recently, on the stress at the onset of instability in compression at low strains ($<$ 10\%) \citep{MLL17, MLL17anisotropy}. Based on these latter simulation strategies, \cite{marvi2018surrogate} built simple and accurate surrogate models for closed-cell foams that can account for the effect of the foam relative density, fraction of mass in struts and cell walls, and cell anisotropy on the elastic modulus and plateau stress.  Nevertheless, the extension of this approach to large strains was limited to open cell foams \citep{Jang20102872} because of the numerical difficulties associated with the development of instabilities in the cell walls and struts, together with the tracking of the pressure within the cells and of contact between struts which leads to the   observed  strain hardening at large strains \citep{MSM09}. Thus, there are no currently high fidelity simulation tools that can establish the link between microstructure and mechanical properties up to large compressive strains (30-70\%) and predict the relevant mechanical properties. They include the effect of cell pressure on the flow stress, the densification strain and the energy absorption and recovery during unloading-reloading cycles. 

In this paper, a strategy to perform high fidelity numerical simulations of the mechanical behavior of closed-cell PU foams under compression up to densification is presented. Digital RVEs of the microstructure are created from the microstructural features (relative density, cell size distribution and shape, fraction of mass in struts and walls) using a methodology previously developed \citep{MLL17, MLL17anisotropy}. The RVEs are discretized using shell elements for the cell walls and beam elements for the struts and the mechanical response under compression is obtained by the finite element simulation of the RVE. The analyses include the effect of the gas pressure within the cells and of the contact and friction between struts and walls during densification. The model predictions were validated against the experimental results in two isotropic and anisotropic closed-cell PU foams. In particular, the model was able to predict the elastic modulus, the stress at the onset of plastic instability, the energy dissipated during deformation and the hardening due to the build up of internal pressure and of the contacts between struts and walls a well as the energy released during unloading. Overall, this new simulation strategy is able to unveil the link between the microstructure and the full set of mechanical properties of closed-cell foams  and, based on the simulation results, a novel analytical model to predict the stress-strain curve of closed-cell foams until densification was proposed. The model can take into account the influence of the most relevant microstructural factors (relative density, cell anisotropy, fraction of mass in cell walls and struts) together with the pressure in the cells on the mechanical response. Finally, as an example application, the simulation strategy was used to design foams with an optimal microstructure for packaging application that require to absorb the maximum energy while the stress transmitted by the foam is below a threshold to avoid damage of the package. 

\section{Simulation strategy}
\label{ComputStrategy}

\subsection{Digital RVE}

A 2 x 2 x 2 mm$^3$ digital twin of the microstructure of the foam was built using the methodology presented in  \citep{MLL17, MLL17anisotropy}. The main steps are summarized below for the sake fo completion. The topology of the foam was obtained by means of the Laguerre tessellation of a cubic domain using NEPER \citep{NEPER}. The Laguerre tessellation is a weighted Voronoi tesselation  in which each seed point is associated to a weight that controls the volume of the cell corresponding to this point. The coordinates and the weight of the seed points of the Laguerre tesselation were obtained from the position of the centres and the radii, respectively, of an ensemble of random close-packed polydisperse spheres, whose size distribution was equal to that of the cell size distribution in the foam as  determined by X-ray microtomography \citep{L2007}. The resulting topology of the foam is isotropicn and anisotropy can be introduced by an affine deformation of all the cells in the Laguerre tesselation by a factor $s$ ($>$ 1) along one of the cube axis, where $s$ stands for cell aspect ratio.

The topology provided by the Laguerre tessellation was discretized using Gmsh \citep{GMsh}. Timoshenko beam elements (type B31 in Abaqus \citep{A16}) and finite-strain shell elements (type S3R in Abaqus) were used to discretize struts and cell walls, respectively. The cross-section of the beam elements was an equilateral triangle  (that is the closest one to the actual three-cusp hypocycloid cross-section of struts in the foams \citep{Jang20081845}) and the area of the beam was a function of the distance to the ends of the strut to reproduce the experimental results on the strut shape obtained by X-ray microtomography \citep{MLL17}. A strut of average length was discretized with approximately 30 beam elements and it was checked that this number was large enough to ensure that the simulation results were independent of the discretization. The shell elements for the cell walls had a constant thickness, which determined the total volume of solid material in the cell walls. The remaining solid material was ascribed to the struts assuming that the strut mass was proportional to its length, in agreement with X-ray microtomography measurements \citep{MLL17}.

\subsection{Numerical simulation}

Numerical simulations of the mechanical behaviour of the RVE of the foam were carried out by solving directly the dynamic equations of motion of the finite element model using Abaqus/Explicit \citep{A16} within the framework of the finite deformations and rotations theory with the initial unstressed state as reference. Previous attempts to carry out the simulations beyond the onset of plastic instability using the implicit finite element method failed because of the buckling of beam and shell elements impaired the convergence \citep{MLL17}. In this work, explicit simulations were performed under quasi-static loading conditions. Mass scaling was used  to increase the explicit stable time integration step. The quasi-static conditions during the simulation were guaranteed by limiting the kinetic energy  to a maximum of 2\% of the total energy of the model and by verifying that the reaction forces on opposite surfaces of the RVE were equal.  The solid PU foam was modelled as an isotropic, elastic-perfectly plastic solid following the J2 theory of plasticity implemented in Abaqus/explicit \cite{A16}.

The effect of the gas pressure inside the closed cells of the foam was introduced in the model using the fluid cavity interaction algorithm in Abaqus/Explicit \citep{A16}. Herein, the term cavity  refers to a set of elements which form a closed surface that remains sealed during the analysis. Each cavity corresponds to a closed cell of the foam. The relationship between the gas pressure $p$ and the volume $V_c$ of the cavity is given by

\begin{equation}
p=\frac{R_gm_cT}{M_a}V_c
\label{eq:3.2}
\end{equation} 

\noindent where $T$ is the absolute temperature (= 299K assuming that compression is isothermal), $R_g$ the universal gas contact (8.314 J/K mol), $M_a$ the average molecular weight of the air within the cells (0.0289 Kg/mol) \citep{rodriguez2009measuring} and $m_c$ the mass of gas in the cell that it is determined from the initial volume of the cell assuming that  the initial pressure in the cavity is $p_0$= 1 atmosphere. 

It is well known that the accuracy of the effective properties obtained by means of the numerical simulation of a RVE of a heterogeneous microstructure increases (for a given RVE size) with the application of periodic boundary conditions  as compared with Dirichlet, Neumann or mixed boundary conditions \citep{SL02, kanit2003determination}. Nevertheless, the implementation of periodic boundary conditions in explicit analysis often leads to spurious displacement oscillations that impairs the numerical analysis \citep{sadaba2019special}. Moreover, the application of periodic boundary conditions requires to develop a periodic microstructure of the foam  and the application of these boundary conditions to beam and shell elements is complicated and requires extra mesh manipulations. However, the discretization of the RVE using beam and shell elements is very efficient from the computational viewpoint and reduces dramatically the computational cost, as compared with solid elements \citep{YMG05}. Thus, it was decided to use mixed-boundary conditions (as detailed below) in combination with larger RVEs containing around 800 cells.

\begin{figure} [t]
  \centering
  \includegraphics[scale=0.4]{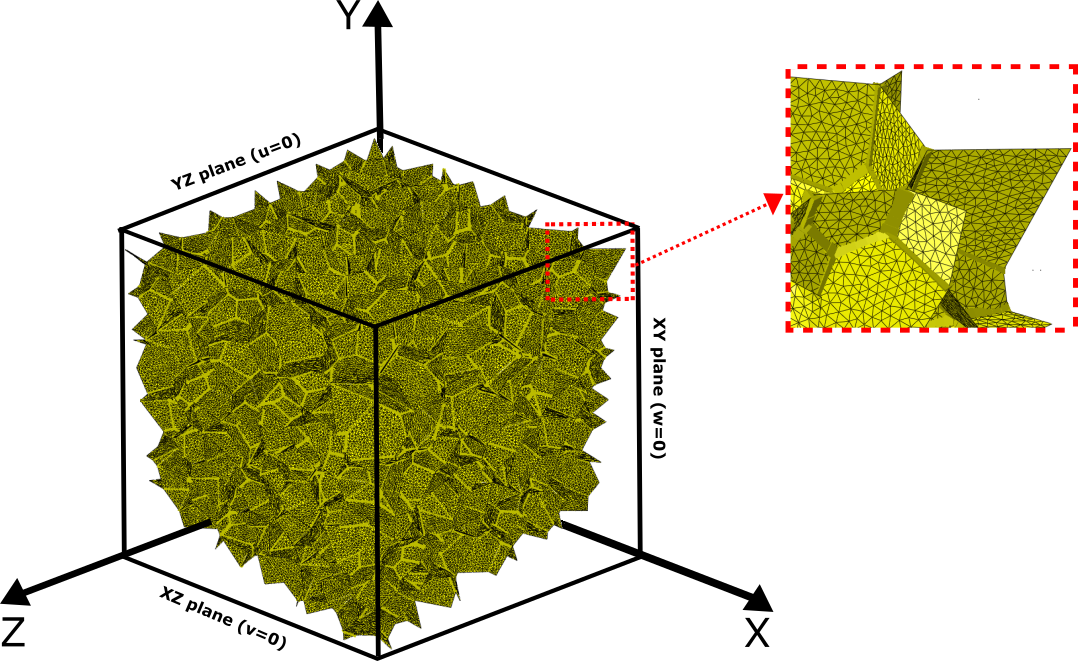}
  \caption{RVE of the foam containing 800 cells together with boundary conditions. The inset shows the details of the finite element discretization of cell walls and struts and the variation of the cross section of the struts along its length can be appreciated.}
  \label{fig:BC}
\end{figure}

The translational degrees of freedom perpendicular  to the planes $XY$, $XZ$ and $XY$ were constrained (Fig. \ref{fig:BC}). If $u$, $v$ and $w$ stand for the displacements along the $X$, $Y$ and $Z$ axes, respectively, $w$ = 0 in the $XY$ plane,  $v$ = 0 in the $XZ$ plane, and $u$ = 0 in the $YZ$ plane. Moreover, all the rotational degrees of freedom were constrained on the  external faces of the RVE. Load was applied by means of the displacement of one rigid surface parallel to one of the faces of the RVE while the opposite face of the RVE was in contact with another rigid surface that was fixed.  The applied strain rate in the simulations was $\approx$ 0.017  s$^{-1}$.  The general contact algorithm in Abaqus  (based on Lagrange multipliers) was used to account for the interaction of the rigid surfaces with the solid material  and  the contacts between the cell walls and struts of the foam upon densification. The contact between the rigid surfaces and the foam was assumed to be frictionless while the friction coefficient  between cell walls and struts was 0.3 assuming Coulombian friction.

\section{Materials and Experimental techniques}
\label{section:ExperimentalResults}

Two PU foams, with isotropic and anisotropic microstructure, were chosen to validate the results of the simulation strategy. The foams, provided by BASF Polyurethanes GmbH (Lemf\"{o}rde, Germany), were obtained from the reaction of a polyol with a polyisocyanate using a mixture of cyclopentane and water as blowing agent. The isotropic and anisotropic foam will be denominated 1-3CPW30 and ACPW73, respectively.

The microstructure of the foams was carefully analyzed in previous investigations \citep{MLL17, MLL17anisotropy}.The cell size distribution in the foams was  determined by means of X-ray  microtomography and  followed a Gaussian distribution characterized by the mean value $\mu$ and the standard deviation $sd$. The variation of the strut cross-section  along the struts  was also determined by means of X-ray computed microtomography and was approximated by a polynomial function, while average thickness of the cell walls ($\overline t$) and the average cell aspect ratio ($s$) were measured from scanning electron microscopy micrographs. The longest cell dimension in the anisotropic  ACPW73 foam was always parallel to the rising direction of foam during manufacturing. The relative density of the foams (assuming a density of the solid PU of 1200 kg/m$^3$) and the microstructural features are summarized in Table \ref{tab:SEM}.

\begin{table}[h]
\centering
\caption{Microstructural features of the PU foams.}
\label{tab:SEM}
\begin{tabular}{lcccc}
\\ \hline
Foam          & 1-3CPW30   & ACPW73  \\
\hline
 Relative density &   0.025  &  0.061 \\
 Average cell size $\mu$ ($\mu$m) &  216  &  354 \\
 Standard deviation of the cell distribution $sd$ ($\mu$m) & 67 & 102 \\
  Average cell aspect ratio $\overline s$   &       $\approx$ 1 &  1.7$\pm$ 0.4\\
Average cell wall thickness $\overline t$ ($\mu$m)    & 0.44 $\pm$ 0.25 &  0.83 $\pm$ 0.05\\
Fraction of solid material in the cell walls $\phi$ & 0.86 & 0.80\\
\hline
\end{tabular}
\end{table}

The elastic modulus, $E_s$, and the yield strength, $\sigma_{ys}$, of the solid PU in the foams were determined from nanoindentation tests, following the methodology presented in \cite{MLL17}. They are depicted in Table \ref{MSP} and were very similar for both foams. However, it was found that the PU in the 1-3CPW30 foam underwent a considerable stiffening with time and the mechanical properties of the PU in this foam after several years of ageing were different from those of the fresh foam. They are also included in Table \ref{MSP} because the mechanical tests involving periodic unloading/reloading cycles in the 1-3CPW30 foam were carried out several years after the monotonic tests. The Poisson's ratio  of the solid dense PU  was assumed to be equal to 0.35 in all cases. Details about the experimental methodology to determine these properties can be found in \citep{MLL17,MLL17anisotropy}.

\begin{table}[]
\centering
\caption{Mechanical properties of solid PU in the foams.}
\label{MSP}
\begin{tabular}{lcc}
\\ \hline
 Foam & $E_s$ & $\sigma_{ys}$ \\ 
 & (GPa) &  (MPa) \\ \hline
1-3CPW30 & 2.4  & 110  \\ 
1-3CPW30 (aged) &  2.7 &  180 \\
ACPW73 & 2.48  & 109  \\ 

\hline
\end{tabular}
\end{table}

The mechanical behavior of the foams in compression until full densification was measured on  cubes of 50 x 50 x 50 mm$^{3}$ machined from the foam panels by wire cutting following the ASTM Standard C297-04. The tests were carried out in a universal electromechanical testing machine  at room temperature under displacement control at a rate of 2 mm per minute, which led to an approximate strain rate of 0.04 s$^{-1}$. The load was measured with a 2 KN load cell, while the average strain was determined from the stroke displacement. Three samples were tested in compression for each foam type and orientation (parallel and perpendicular to the longest cell dimension) in the case of the anisotropic foam. In addition, another set of mechanical tests was carried out for each foam and orientation with periodic unloading/reloading cycles to determine the amount of energy stored and dissipated as a function of the applied strain. 
  
\section{Simulation results and discussion}

The simulation strategy presented above was used to simulate the mechanical response until full densification of the isotropic and anisotropic foams, the latter in the orientations parallel and perpendicular to the rising direction of the foam. The microstructure and mechanical properties of the foams to build the digital RVE are given in Tables \ref{tab:SEM} and \ref{MSP}. 
The effects of the gas pressure and friction during crushing on the mechanical response of the foam are first analyzed in section \ref{mech} taking the isotropic foam as example. Then, the simulation results for RVEs containing 100 and 800 cells were carried out and compared with the experimental results for the case of isotropic and anisotropic foams in sections \ref{iso} and \ref{aniso}  while the effect of gas pressure in the cells  and of the estimations of the energy stored and dissipated during deformation are discussed in sections \ref{gp} and \ref{un-re}, respectively. They are also compared with the experimental results.

\subsection{Effect of gas pressure and friction during crushing} \label{mech}

\begin{figure} [t!]
  \centering
  \includegraphics[scale=0.5]{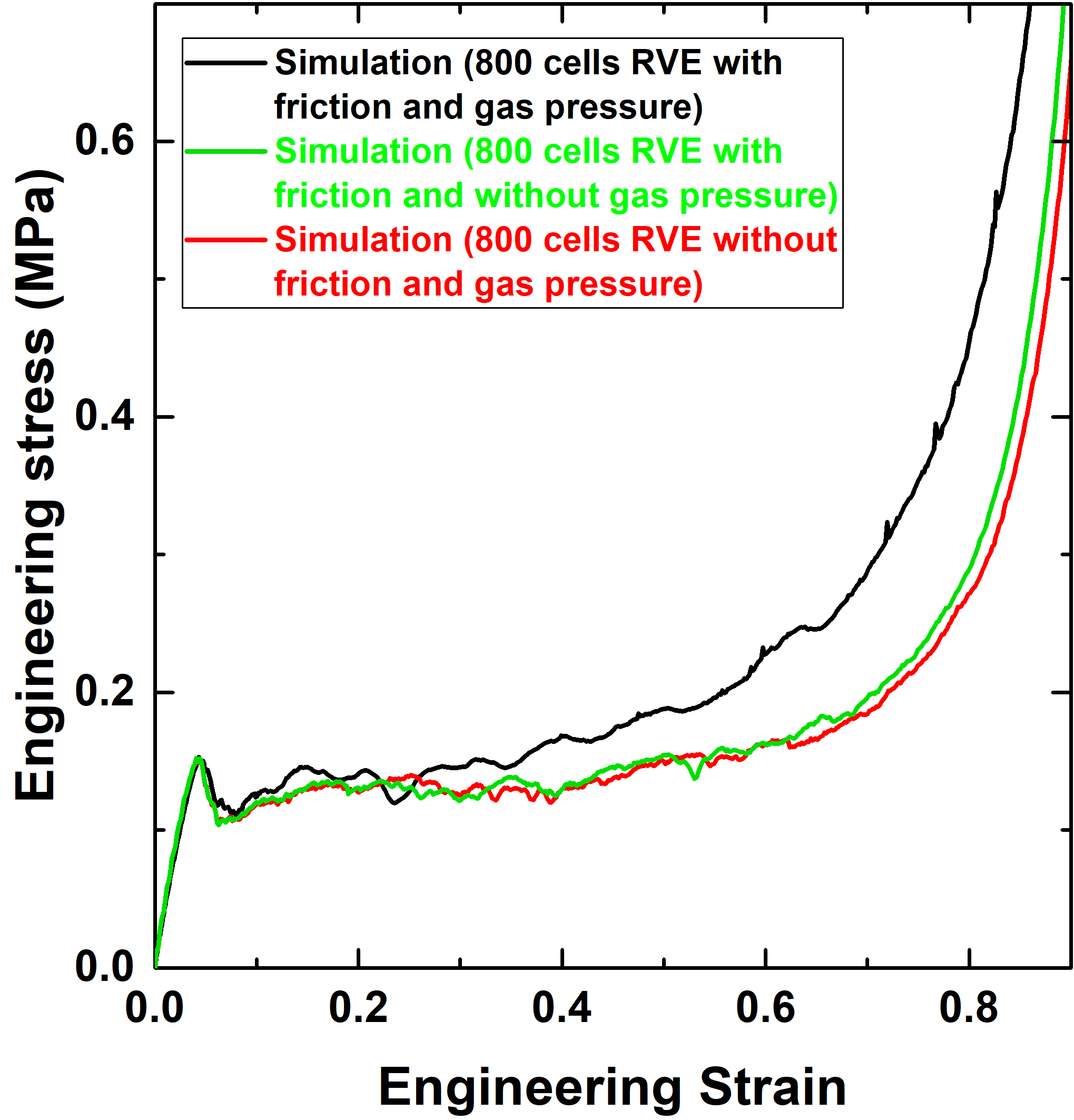}
  \caption{Simulated engineering stress-strain curves in compression of the isotropic 1-3CPW30 foam considering different deformation mechanisms. All the simulations were performed in the same RVE with 800 cells.}
  \label{fig:mech}
\end{figure}

The simulated engineering stress - engineering strain curves obtained with an RVE including 800 cells are plotted in Fig.~\ref{fig:mech} for the isotropic foam considering different effects: neither friction nor gas pressure, friction but not gas pressure and friction plus gas pressure. The results of these three simulations are practically superposed in the elastic regime and at the beginning of the plateau region (strain $<$ 30\%) because neither friction nor gas pressure are important at these deformations. However, the influence of the gas pressure within the cells has a strong influence on the load bearing capability of the foams for larger strains and only the simulations that included the effect of the gas pressure were able to reproduce accurately the hardening of foam in this regime. In particular, the simulations including the effect of gas pressure showed that the strain hardening began to increase rapidly for compressive strains $\approx$ 50\% while this effect was delayed until 75\% in the simulations that did not consider the gas pressure. The influence of the friction on the mechanical response was only important in the last stages of densification (strain $>$ 80\%) but it should be noted that all the simulated curves tended to collapse in this regime. Moreover,  the discretization with beam and shell elements is not the most appropriate to reproduce the foam behavior when the relative density has increased dramatically due to compaction and it can be concluded that neglecting the effect of friction will not affect the accuracy of the simulations. Nevertheless, they were included in the simulations below because the additional computational cost was negligible.

\subsection{Isotropic foam} \label{iso}

\begin{figure} [t!]
  \centering
  \includegraphics[scale=0.5]{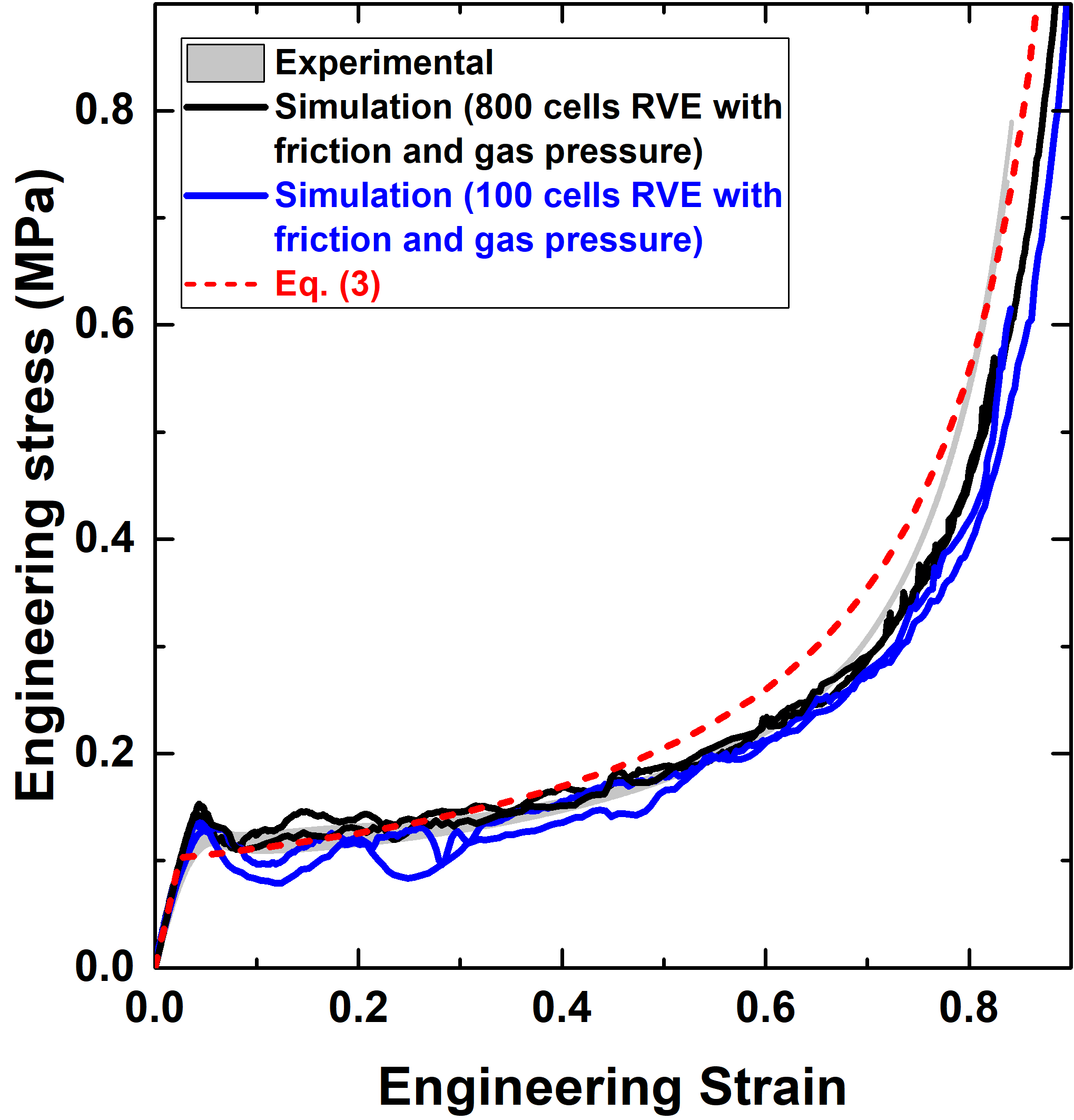}
  \caption{Experimental and simulated engineering stress-strain curves in compression of the isotropic 1-3CPW30 foam. The shaded grey area encompasses the experimental results. Simulations were carried out in RVEs containing either 100 and 800 cells (2 simulations for each). In addition, the predictions of the analytical model, eq. \eqref{Analytical}, are plotted as a dashed curve.}
  \label{fig:IsoStressStrain}
\end{figure}

The simulated and experimental engineering stress-strain curves of the 1-3CPW30 foam are plotted in Fig.~\ref{fig:IsoStressStrain}. The experimental results are plotted as a shaded grey area that includes the three experimental curves while 4 curves corresponding to different numerical simulations are plotted in the figure. Two of them were carried out with 800 cells in the RVE and another two with only 100 cells. The effects of gas pressure and friction were included in all simulations. 

The simulated stress-strain curves with 100 and 800 cells were practically superposed up to the onset of the non-linear region. Afterwards, the differences were small although the curves obtained with 100 cells were slightly softer and showed several stress drops in the plateau region. They were linked to the simultaneous collapse of several cells during deformation, which has a strong influence in the stress carried by the RVE when the number of cells is small. In a previous investigation \citep{MLL17}, we reported that RVEs containing 100 cells were large enough to predict accurately the elastic modulus and the plateau stress of the foam and this conclusion can be extrapolated to the whole stress-strain curve in the case of isotropic foams although the instabilities were more marked in the simulations with 100 cells.

The simulation results were in good agreement with the experimental curves along most of the stress-strain  curve and captured accurately the progressive hardening of the foam due to the build-up of pressure within the cells, which was particularly important for applied strains $>$ 40\%. The agreement with experiments in the final densification region was poorer for the reasons indicated at the end of the previous section.

\begin{figure} [!]
  \centering
  \includegraphics[scale=0.40]{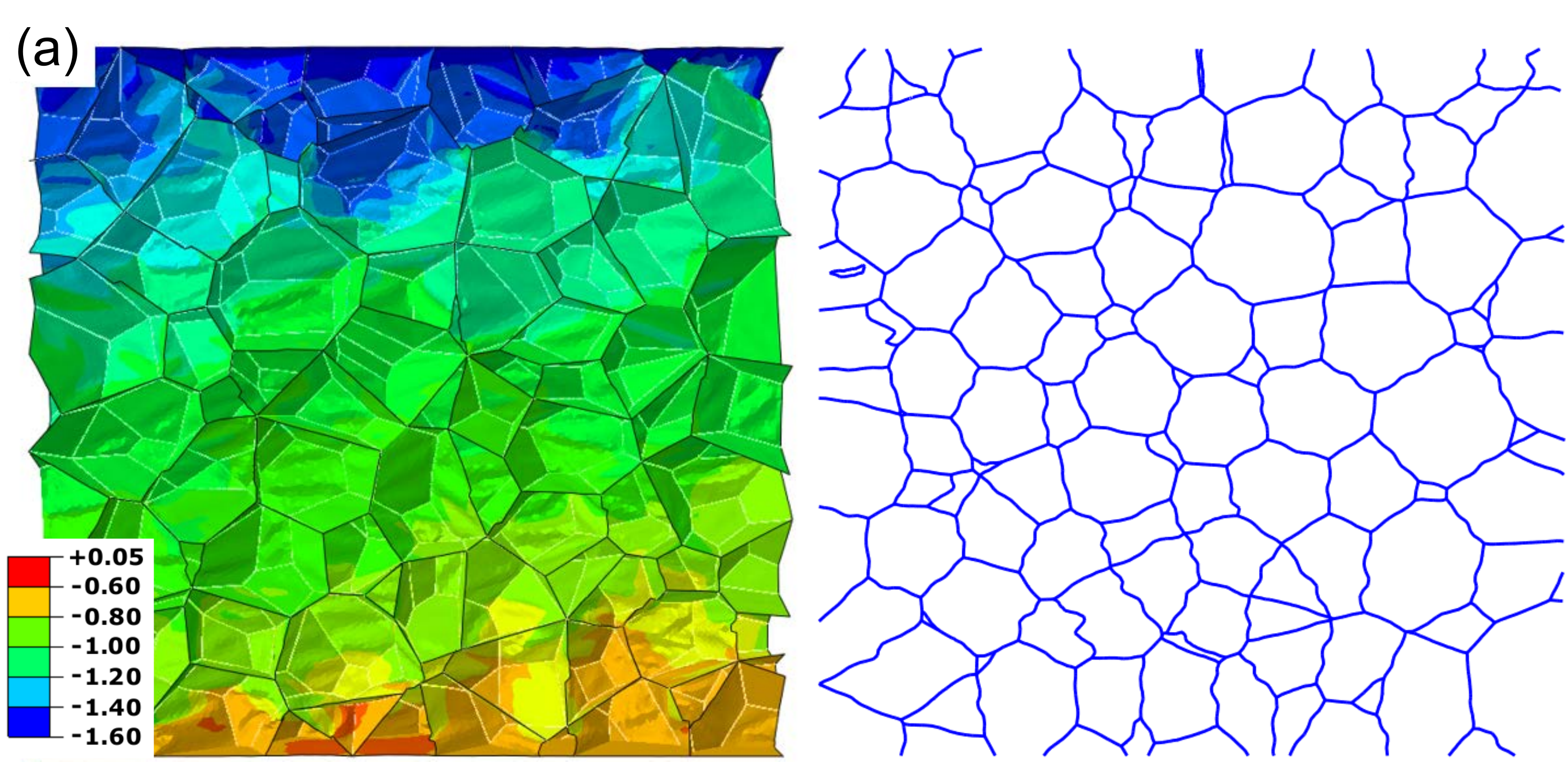}
  \includegraphics[scale=0.40]{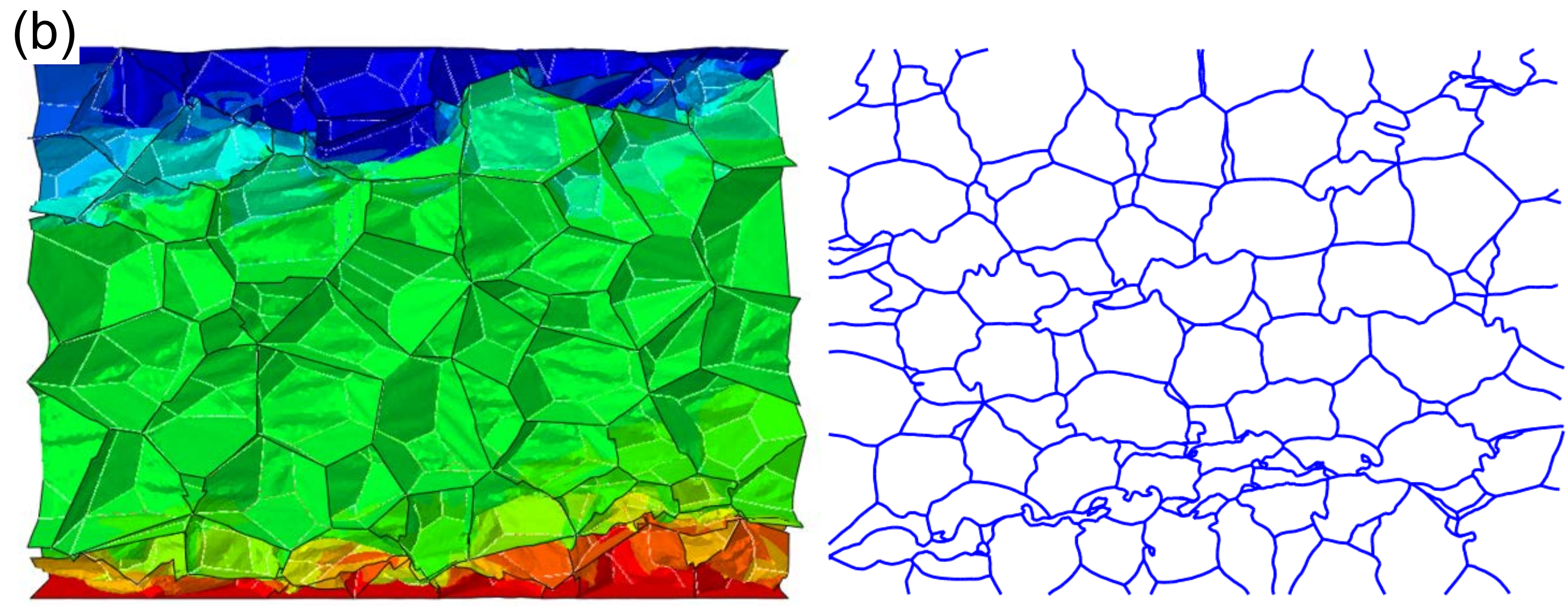}
  \includegraphics[scale=0.40]{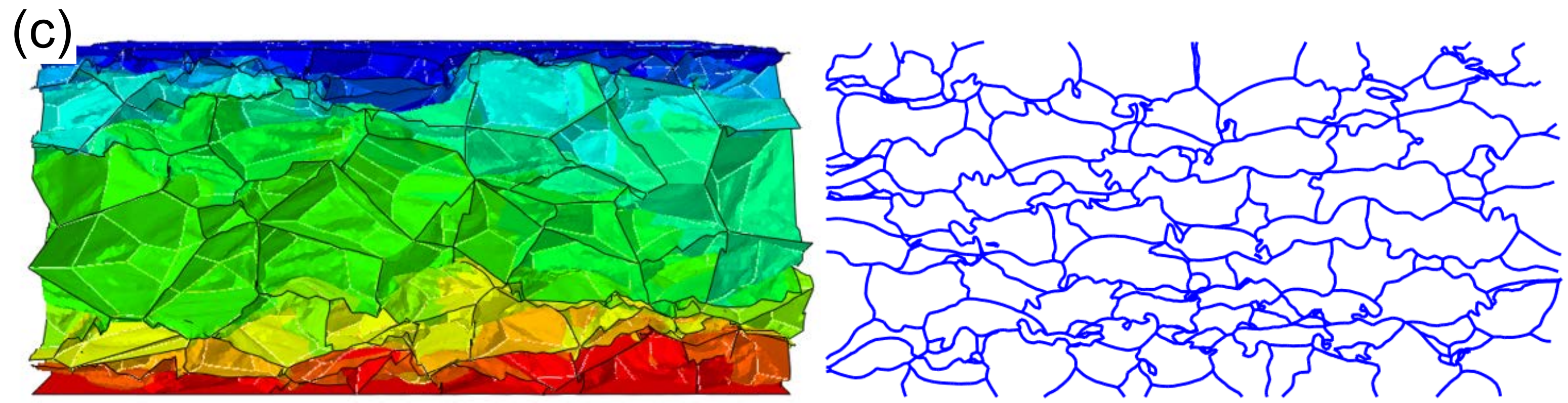}
  \includegraphics[scale=0.40]{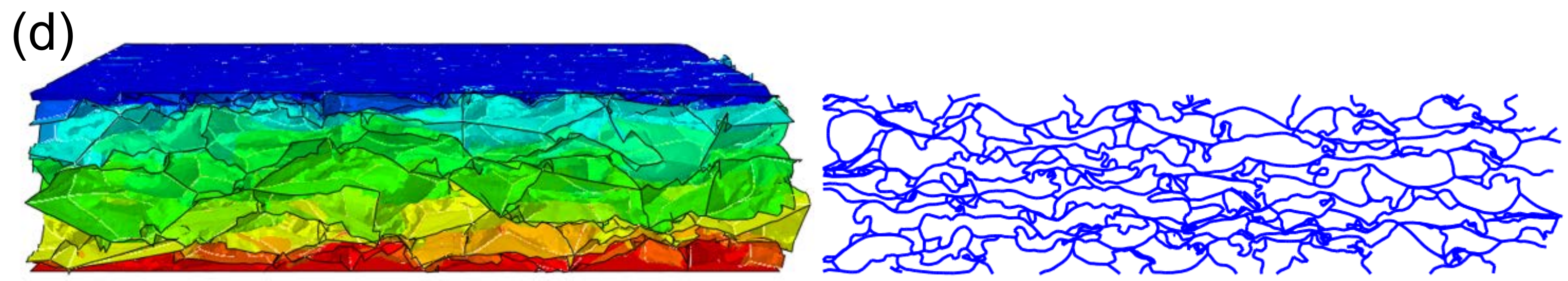}
  \caption{Contour plots of the displacement in vertical (loading) direction on one of the lateral surfaces of the foam and and deformation of the struts in the central section of the foam. (a) Compressive strain of 4\%. (b) Compressive strain of 25\%. (c) Compressive strain of 50\%. (d) Compressive strain of 75\%. The RVE contained 800 cells and the simulations included the effect of friction and gas pressure. The displacement in the legend is expressed in mm and the initial height of the RVE was 2 mm.}
  \label{fig:deformation}
\end{figure}

The deformation mechanisms of the isotropic foam are depicted in Fig. \ref{fig:deformation} as a function of the applied compressive strain. The left column of the figure shows the contour plot of vertical displacement (parallel to the loading direction) on one of the lateral surfaces of the RVE while the right column shows the deformation of the struts in the central section of the RVE. The deformation in the elastic region is homogeneous (Fig. \ref{fig:deformation}a) but there is evidence of buckling in the cell walls and in the struts even in the elastic regime. Localization of deformation in one section of the RVE is evident in the image of the deformation of the struts when the deformation has reached 25\% (Fig. \ref{fig:deformation}b), indicating that the collapse of the foam structure takes place sequentially. Several collapsed layers can be observed when the deformation has reached 50\% (Fig. \ref{fig:deformation}c) and the densification process is evident at 75\% compressive strain (Fig. \ref{fig:deformation}d). These observations are in very good agreement with the experimental observations of the compressive behavior of closed-cell foams \citep{LSG05, ABI07, OA09, TEKOGLU2011109, MLL17}.

\subsection{Anisotropic foam}\label{aniso}
\begin{figure} [t]
  \centering
  \includegraphics[height=6.9cm]{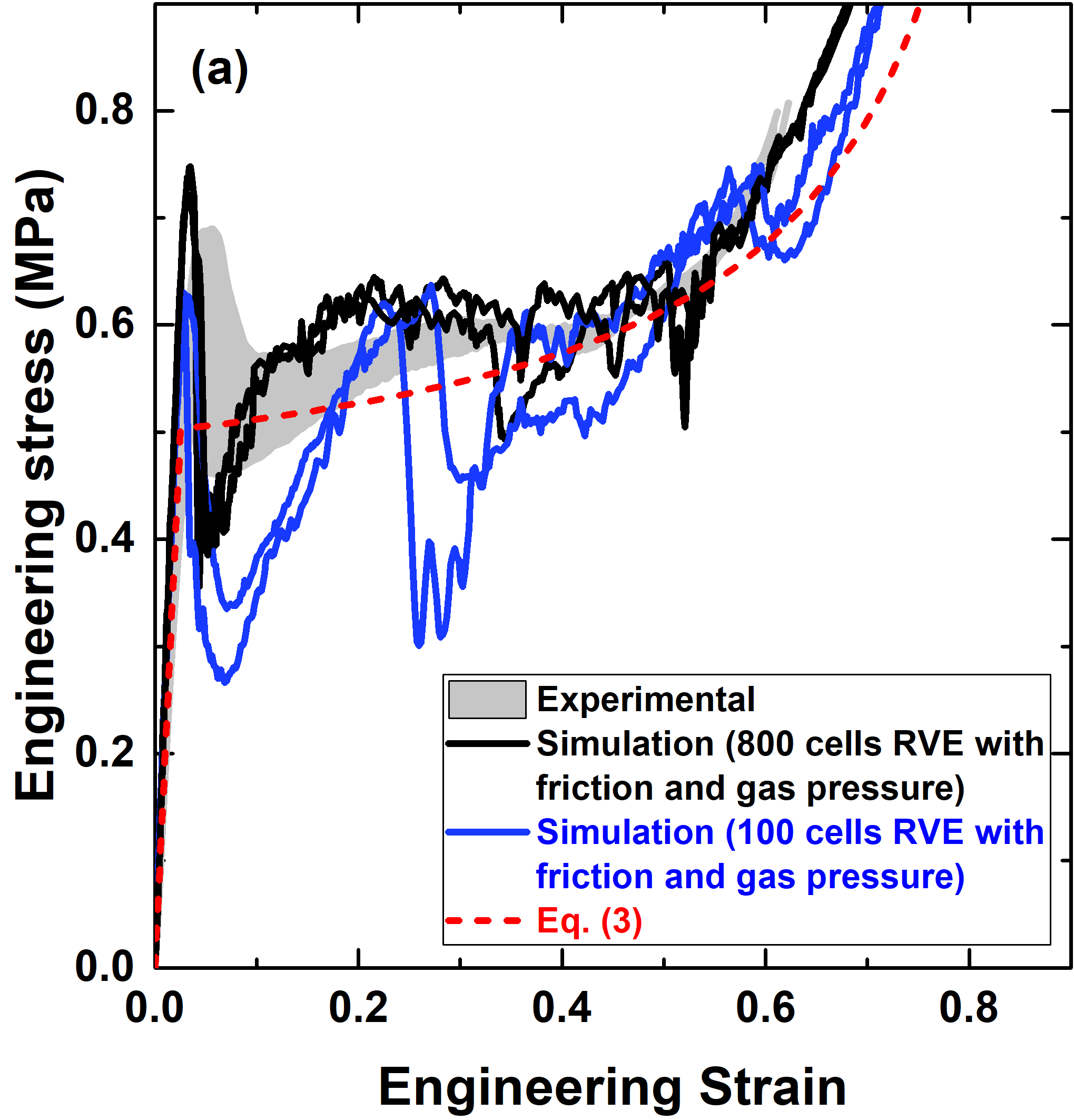}
   \includegraphics[height=6.9cm]{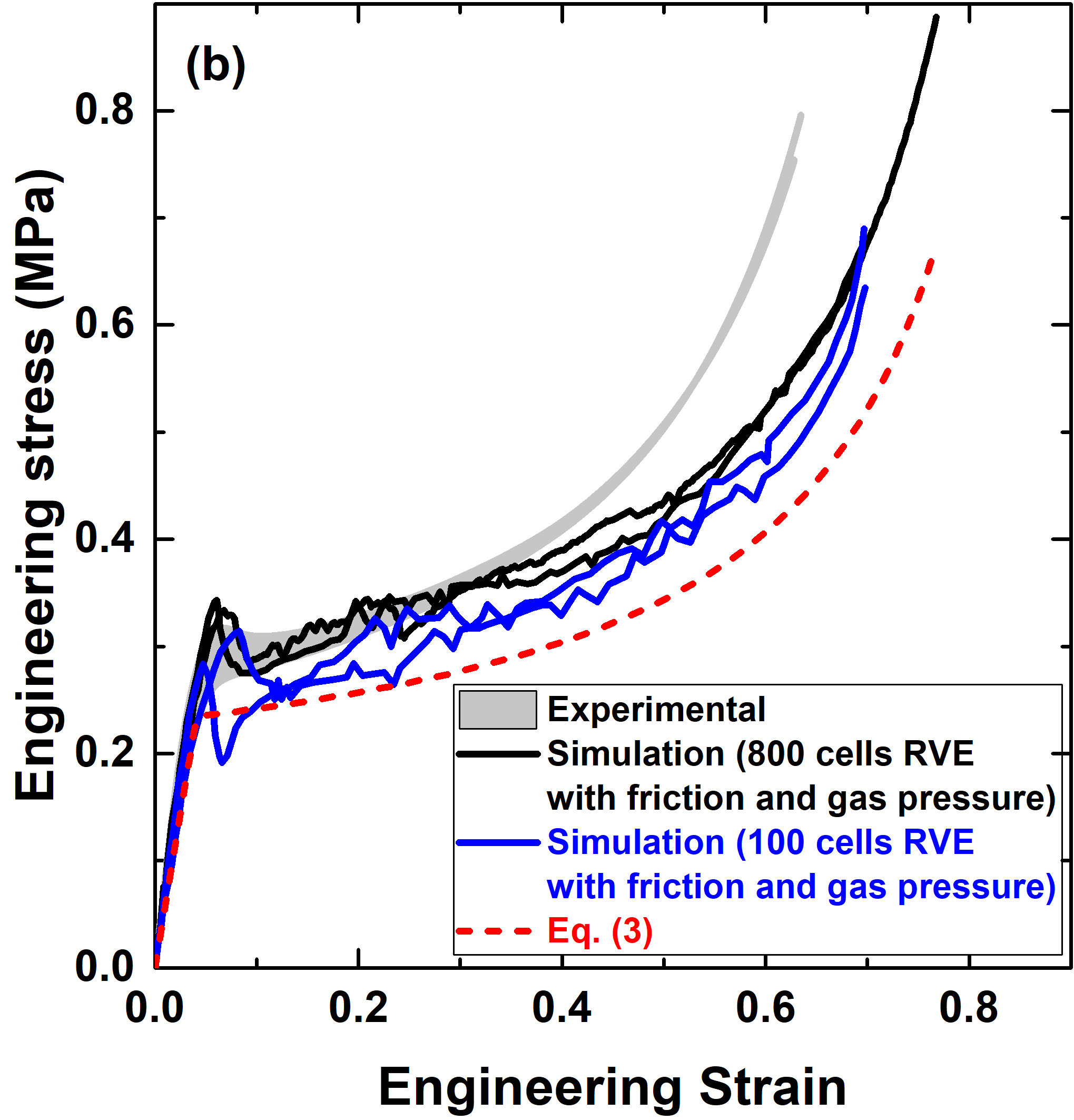}
  \caption{Experimental and simulated engineering stress-strain curves in compression of the anisotropic ACPW73 foam.  (a) Deformation parallel to the longest axis of the cells. (b) Deformation perpendicular to the longest axis of the cells. The shaded grey areas encompass the experimental results. In addition, the predictions of the analytical model, eq. \eqref{Analytical}, are plotted as a dashed curves.}
  \label{fig:AnisoStressStrain}
\end{figure}

Two digital RVEs with 100 cells and another two with 800 cells were created using the parameters of the ACPW73 foam in Tables \ref{tab:SEM} and \ref{MSP}.  These RVEs were isotropic and the anisotropy was introduced by means of an affine deformation of all the nodes in the RVE by a factor $s$ = 1.7 parallel to one direction ($Y$ axis), according to the experimental value of the anisotropy in Table~\ref{tab:SEM}. Simulations were carried out in the orientations parallel and perpendicular to the longest axis of the cells and the corresponding engineering stress-strain curves in both orientations are plotted in Fig. ~\ref{fig:AnisoStressStrain} together with the corresponding experimental results, which are shown as shaded areas that include all the experimental results. All the simulations included the effects of gas pressure and friction.  The results of the simulations carried out with 100 cells were superposed to those obtained with 800 cells in the elastic region and also predicted similar values of the stress at the onset of plastic instability, in agreement with previous analyses \citep{MLL17anisotropy}. However, they predicted lower stresses in the plateaus region and showed very large oscillations in the stress  when the foams were deformed along the longest axis of the cells as a result of the simultaneous collapse of several cells during deformation. This effect is more noticeable in this orientation than in the perpendicular orientation because the stresses carried out by the cells are higher in the former. Thus, RVEs containing 800 cells were necessary to reproduce accurately the mechanical response in compression of the anisotropic foams deformed along the longest axis of the cells and the comparison with the experimental data below was carried out for these RVEs.

The simulations with 800 cells were able to reproduce the three different regions in the stress-strain curves of the anisotropic foam, namely the elastic regime, the plateau region and the densification region, as well as the large differences due to the anisotropy of the foam. Compression in the parallel orientation led to large increase in the elastic modulus of the foam and in the stress at the onset of plastic instability, that was followed by a marked load drop, in excellent agreement with the experimental results. This was followed by  the plateau region at which deformation progressed at constant stress. The simulated curves  were also able to reproduce the experimental plateau stress in this region. Finally, the onset of densification at strains $>$ 60\% was also  captured by the simulations. 

The simulations perpendicular to the longest axis of the cells were also in agreement with the experimental results and captured the reduction in stiffness and in the stress at the onset of plastic instability, as compared with the parallel orientation. Moreover, no stress drop was found in either the simulations or experiments after this point and both curves showed constant hardening without a well-defined plateau region. Nevertheless, the hardening in this region was underestimated by the simulations. The experimental curves for both orientations met at an applied strain of 60\% but the stress carried by the foam in the simulations in the perpendicular direction was significantly lower than that in the parallel orientation at this strain.

The differences in the initial stiffness and in the stress at the onset of instability between both orientations were analyzed previously \citep{MLL17anisotropy}. They were attributed to a change in the deformation mechanisms, which were  dominated by the axial deformation of the struts in the parallel orientation while bending controlled the deformation perpendicular to longest cell dimension. The stress at the onset of plastic instability was controlled by the buckling of the struts, which was triggered at lower stresses in the direction perpendicular to the rising direction because of bending. Nevertheless, buckling of the struts in the parallel orientation at higher stresses led to the localization of the deformation in one layer of cells in the foam and to the sudden load drop. The simulations in Fig. \ref{fig:AnisoStressStrain} show that further deformation in the parallel orientation led to a rising of the load until the collapse of another layers of cells and this process was repeated several times before densification started when the applied strain reached  60\%. On the contrary, deformation along the perpendicular direction was similar to the one reported in the isotropic foam and the continuous strain hardening was mainly attributed to the effect of the gas pressure within the cells.

\subsection{Influence of the gas pressure on the mechanical behavior}\label{gp}

\begin{figure} [h!]
  \centering
  \includegraphics[scale=0.8]{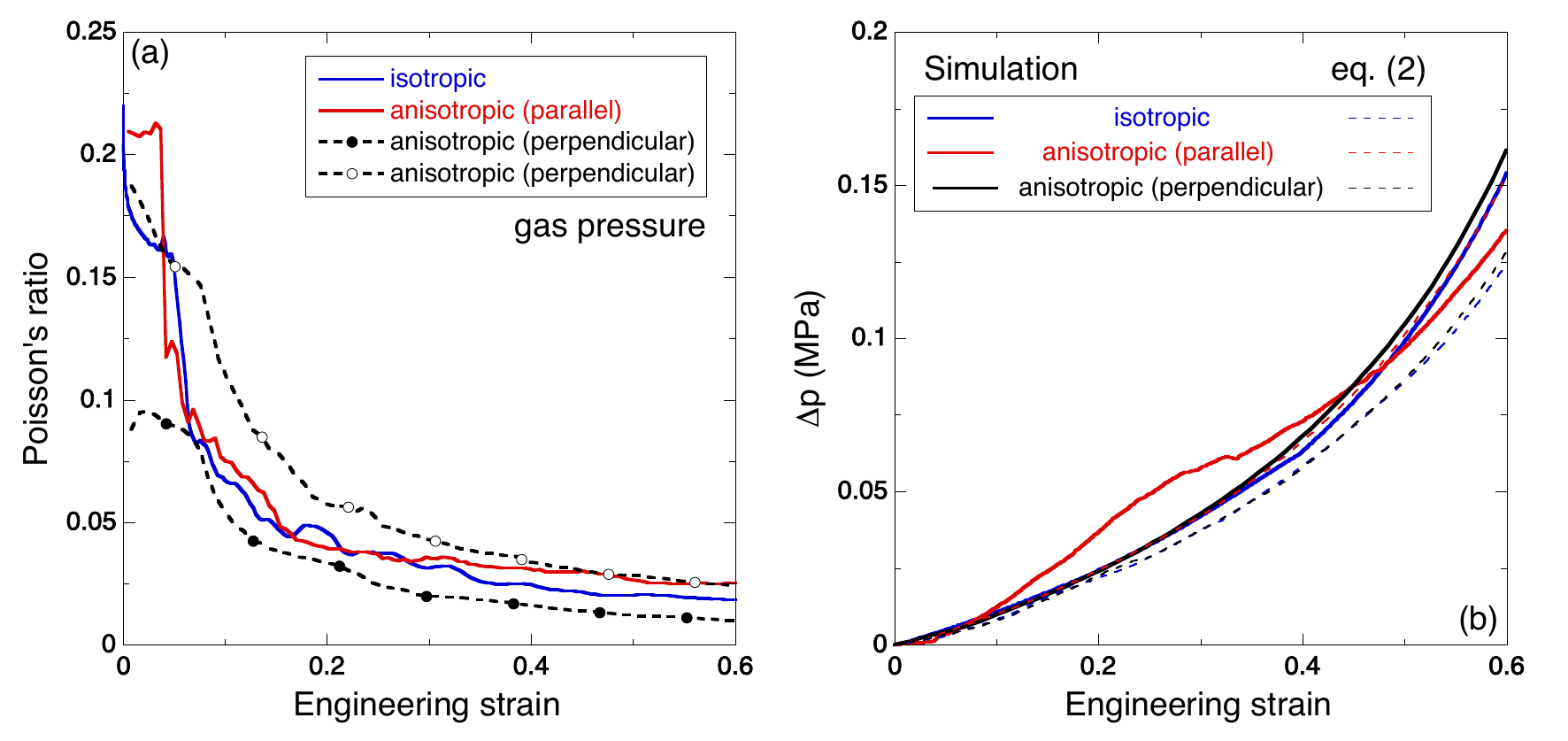}
    \includegraphics[scale=0.8]{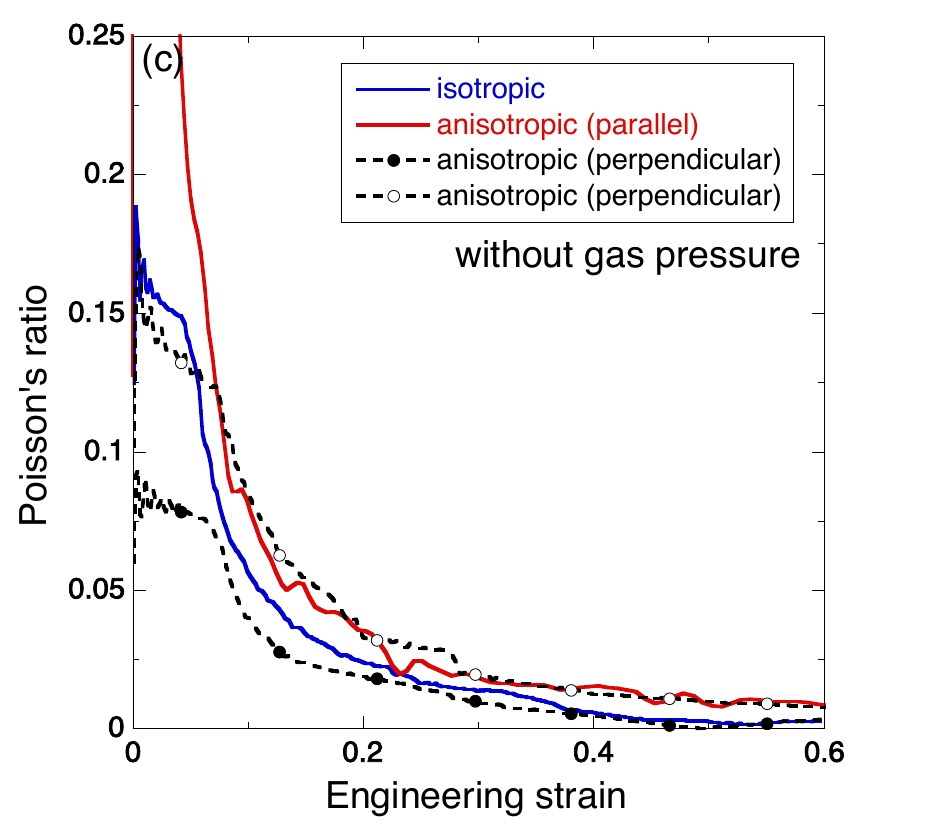}
  \caption{(a) Evolution of the Poisson's ratio of the isotropic and anisotropic foams (the latter deformed in the parallel and perpendicular orientations) as a function of the applied strain. (b) Evolution of the average gas pressure in the cells, $p-p_0$, in the isotropic and anisotropic foams (the latter deformed in the parallel and perpendicular orientations) as a function of the applied strain. The predictions of eq. \eqref{eq:pres} are also plotted for comparison. (c) Evolution of the Poisson's ratio of the isotropic and anisotropic foams (the latter deformed in the parallel and perpendicular orientations) as a function of the applied strain if the effect of gas pressure is not considered.}
  \label{fig:NU}
\end{figure}

The effect of the gas pressure on the mechanical response in compression of the closed-cell foams is evident in the stress-strain curves in Fig. \ref{fig:mech}. This process is closely related to the reduction in the foam volume during compression, which in turns depends on the Poisson's ratio of the foam. In most elasto-plastic solids, the Poisson's ratio in the elastic regime is in the range 0.2-0.35 and increases up to 0.5 (incompressible solid) in the fully plastic regime. The  Poisson's ratio of the foams was determined  in the simulations of the RVEs from the average lateral deformation in both orientations perpendicular to the load axis divided by the applied strain. Its evolution as a function of the applied strain is plotted in Fig. \ref{fig:NU}(a) for the isotropic and anisotropic foams (the latter for deformations parallel and perpendicular to the longest axes of the cells). In the case of the anisotropic foam deformed perpendicular to the longest axis of the cells, two values of the Poisson's ratio, $\nu_{yx}$ and $\nu_{zx}$ are plotted in the figure. They were different because of the different aspect ratios of the cells along the $Y$ and $Z$ axes. 

The initial value of the Poisson's ratio of the foams in the elastic regime was $\approx$ 0.2 for the isotropic foam and the anisotropic foam deformed in the parallel orientation and slightly lower for the anisotropic foam in the perpendicular orientation. It decreased rapidly in the plateau region in all cases to reach a minimum value of $\approx$ 0.02 when the applied strain reached 60\%.  The low value of the Poisson's ratio is associated with a large reduction of the foam volume during compression and to an increase of the gas pressure within the cells, $p$. Under isothermal conditions, the relationship between the gas pressure and the cell volume is given  by eq. \eqref{eq:3.2}. The reduction of the volume during compression is given by volumetric strain $\epsilon_v = \epsilon [1-2\nu(\epsilon)]$, where  $\nu$ is the Poisson's ratio of the foam  that depends on the applied strain $\epsilon$. In the case of deformation perpendicular to the longest axis of the cells in the anisotropic foam, $\nu$ = $(\nu_{yx}+ \nu_{zx})/2$. Thus, the increase of the gas pressure, $\Delta p$, can be estimated as a function of the applied strain $\epsilon$ according to \citep{GibsonBookCellular}

\begin{equation}
\Delta p = \frac{p_0(1-\rho/\rho_s)}{(1-\rho/\rho_s) - \epsilon [1-2\nu(\epsilon)]} -p_0
\label{eq:pres}
\end{equation} 

\noindent where $p_0$ is the initial atmospheric pressure and $\rho$ and $\rho_s$ stand for the density of the foam and of the solid material in the foam, respectively.  The predictions of the cell pressure in the foams according to eq. \eqref{eq:pres} are compared with the average values obtained in the numerical simulations in Fig. \ref{fig:NU}(b) and they are in reasonable agreement. It should be noted, however, that the application of eq. \eqref{eq:pres} will grossly underestimate the gas pressure if the constant value of the Poisson's ratio of the foam in the elastic regime is used.

The simulation results for the Poisson's ratio of the isotropic and anisotropic foams without including the effect of gas pressure have been plotted in Fig. \ref{fig:NU}(c). The general trends are similar to those observed in Fig. \ref{fig:NU}(a), when gas pressure is considered, but the Poisson's ratio in the plateau region is closer to 0 if gas pressure is not considered in the simulations. This behavior is reasonable as the gas pressure opposes the applied strain and enhances the deformation in the perpendicular directions.

\subsection{Energies stored and dissipated during deformation}\label{un-re}

\begin{figure} [t!]
  \centering
  \includegraphics[scale=0.35]{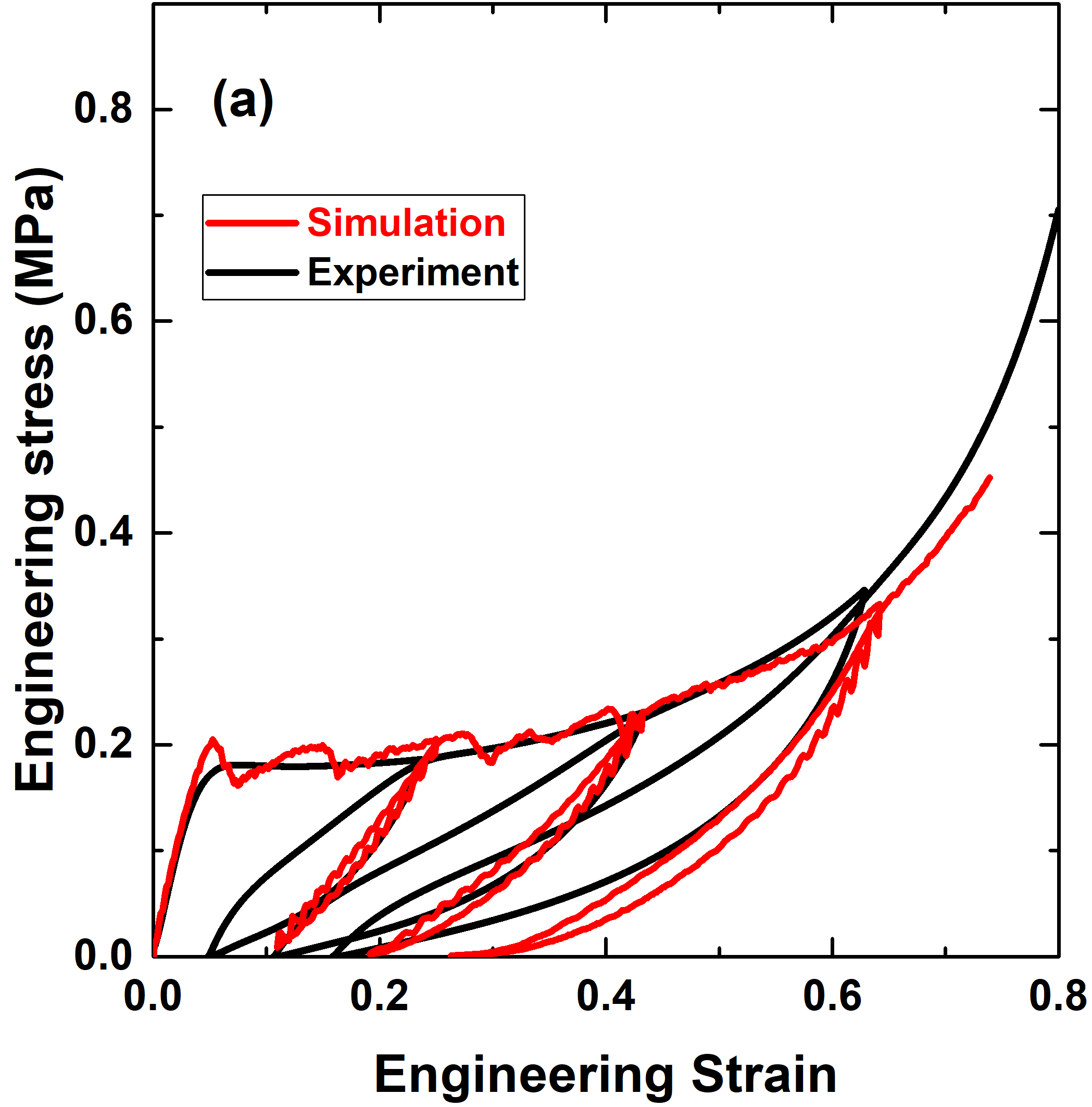}
  \includegraphics[scale=0.35]{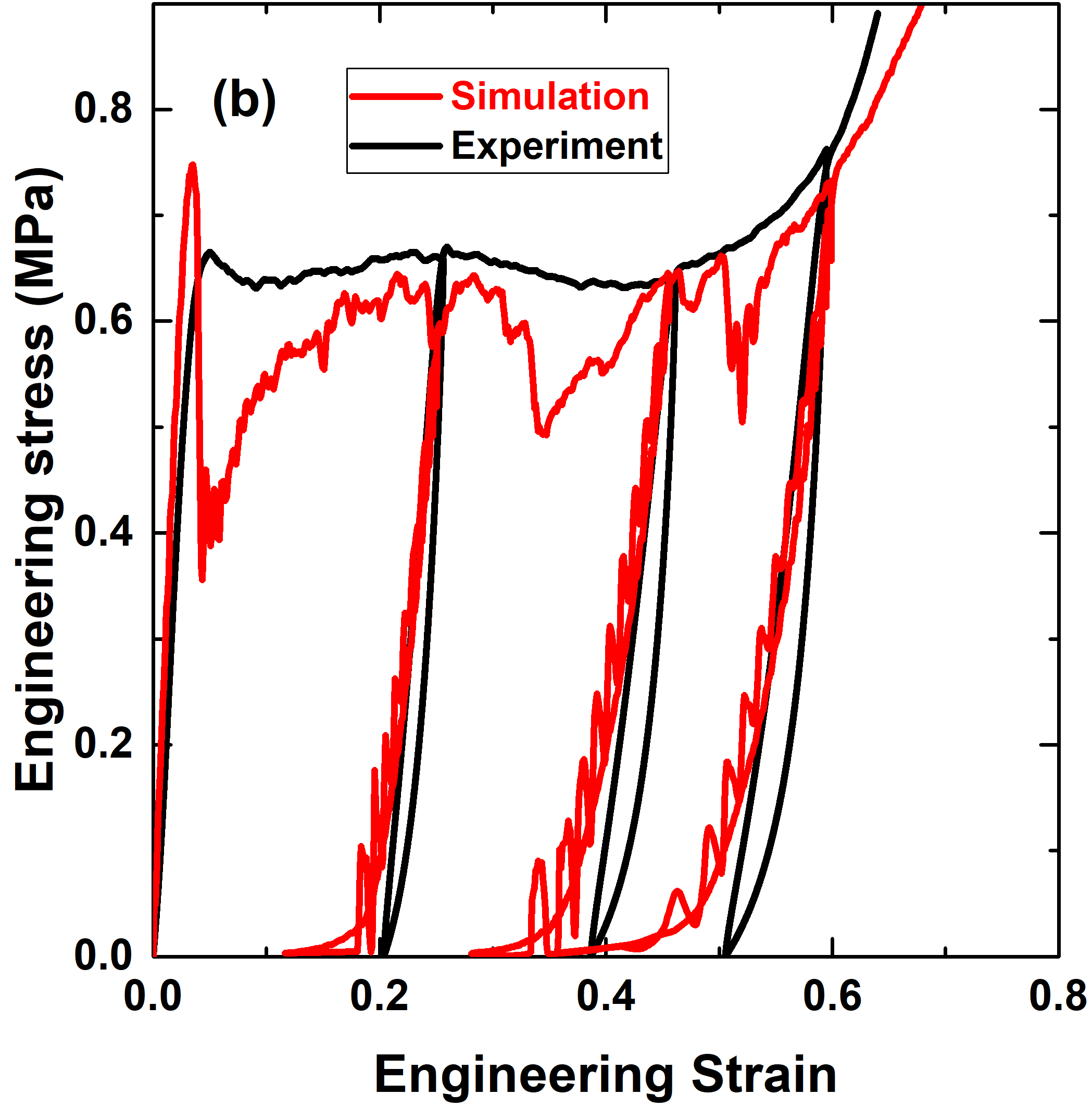}
  \includegraphics[scale=0.35]{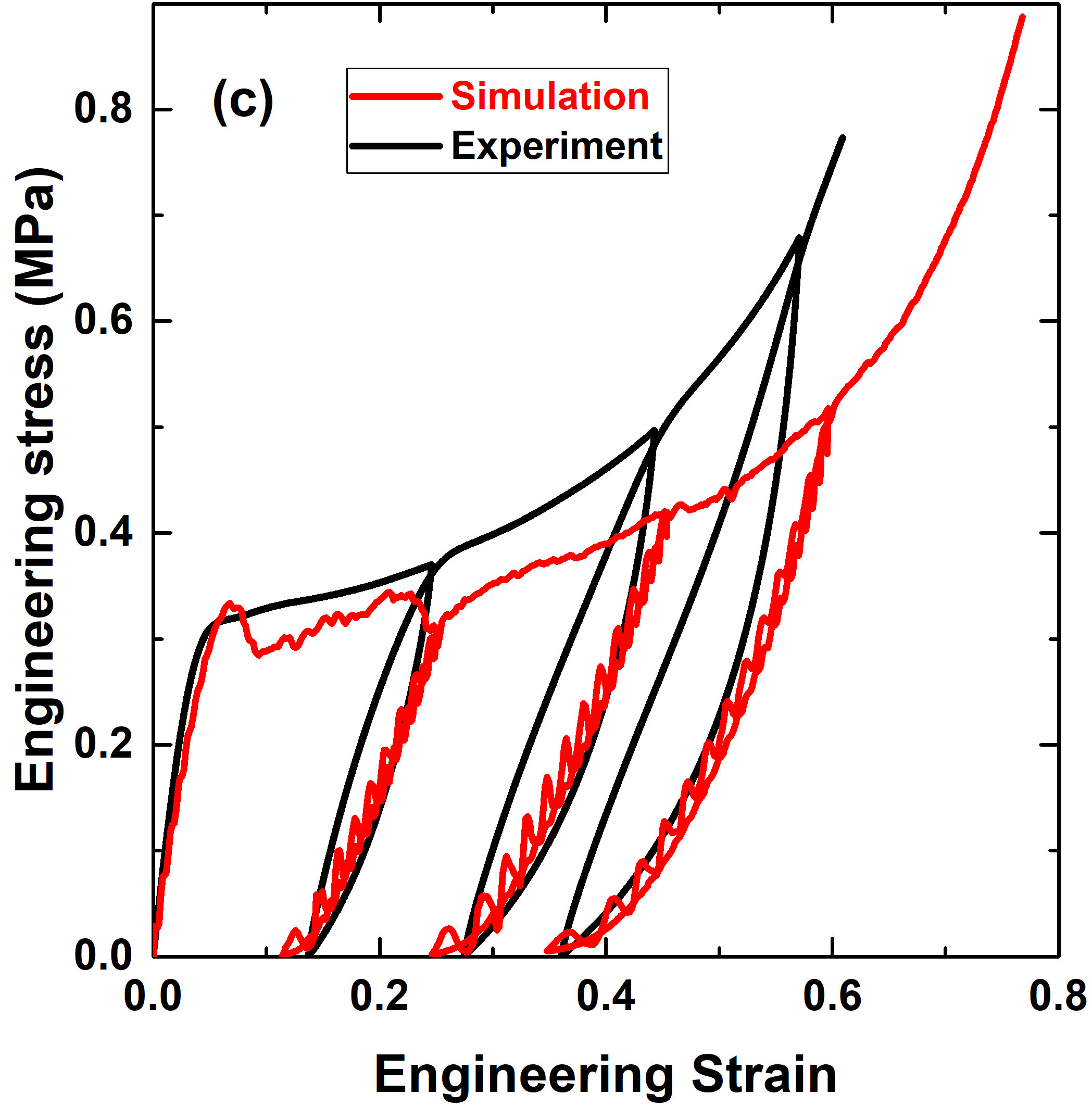}
  \caption{Experimental and simulated engineering stress-strain curves in compression of the foams subjected to periodic unloading/reloading cycles. (a) Isotropic 1-3CPW30 foam. (b) Anisotropic ACPW73 foam deformed along the longest axis of the cells. (c)  Anisotropic ACPW73 foam deformed in the direction perpendicular to the longest axis of the cells.}
  \label{fig:load-unload}
\end{figure}

In order to estimate the energy stored and dissipated as a function of the applied strain, new compression tests were carried out in the isotropic and anisotropic foams in which they were subjected to an unloading/reloading cycle when the applied strain reached 25\%, 45\% and 60-65\%.  The  experimental engineering stress-strain curves are plotted in Fig. \ref{fig:load-unload} together with the results of the simulations in RVEs containing 800 cells and including the effect of gas pressure and friction. The mechanical properties of the solid PU in the simulations of the 1-3CPW30 foam correspond to those denominated aged in Table \ref{MSP}.
The numerical simulations were able to capture the overall shape of the experimental stress-strain curves and also provided a good estimation of the  residual strain after unloading for different values of the applied strain in the isotropic and anisotropic foams. In the case of the anisotropic foams, the stress carried by the foam in the plateau region was slightly underestimated by the simulations. This difference may be due to a slight aging the solid PU in the 1-3CPW30 foam, as the mechanical tests with periodic unloading/reloading were carried out three years after the monotonic tests. However, the aging of the solid PU in the isotropic ACPW73 foam was very limited and the same properties for the solid PU were used in the simulations in Figs. \ref{fig:IsoStressStrain} and \ref{fig:load-unload}(a).

The experimental unloading/reloading cycles showed a large hysteresis in the case of isotropic foam. This hysteresis was reduced in the case of the anisotropic foam deformed perpendicular to the longest axis of the cells and was minimum for deformation along the longest cell axis. The hysteresis was always associated with the non-linear response of the foam during unloading (which was maximum in the case of the isotropic foam and minimum in the anisotropic foam deformed parallel to the longest cell axis) because the re-loading segments tended to be linear in all cases. The non-linearity during unloading was  well captured by the numerical simulations but re-loading in the simulations was also non-linear, following a path very close to the unloading curve and the simulated hysteresis was always very small. Moreover, the re-loading curves in the anisotropic foams showed a large number of oscillations that were attributed to dynamic effects. In fact, they were reduced if the strain rate in the simulations was decreased though this made the simulations impractical. The differences in the re-loading curves between simulation and experiments (particularly in the case of the isotropic foam and of the anisotropic foam deformed in the perpendicular direction) were attributed to viscoelastic effects of the PU, that were not included in the model.

 \begin{figure} [h!]
  \centering
  \includegraphics[scale=1]{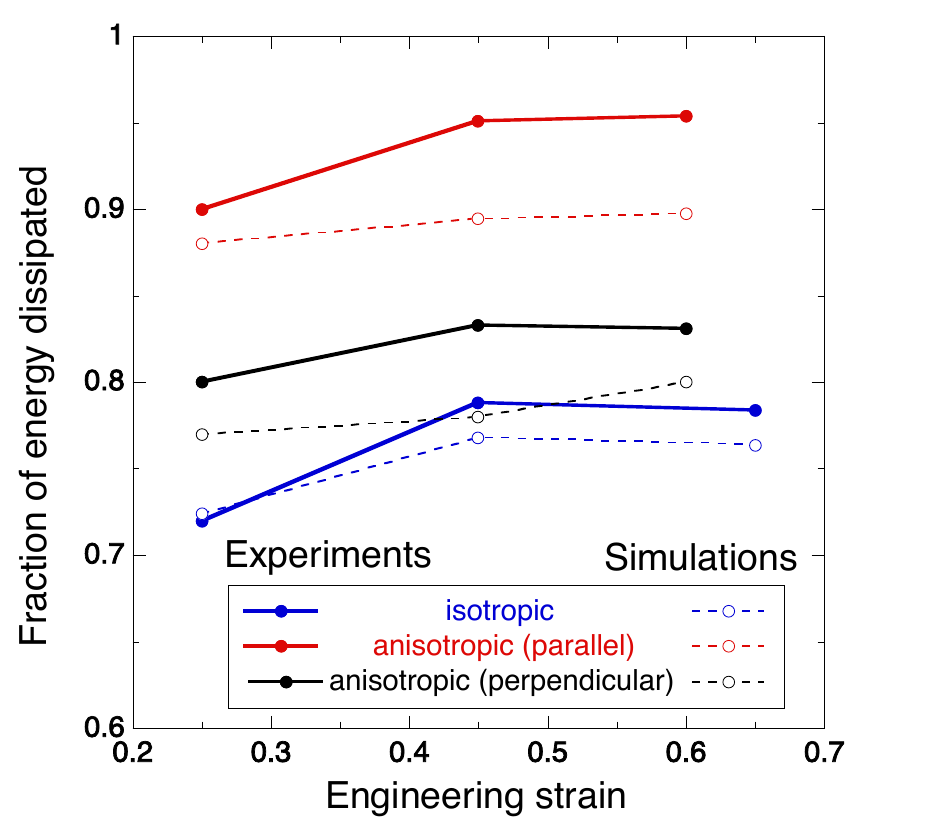}
  \caption{Fraction of energy dissipated as a function of the applied strain obtained from the compression experiments (solid symbols) and simulations (open symbols) including periodic unloading/reloading cycles.}
  \label{fig:DissipatedE}
\end{figure}

The elastic energy stored in the foam for a given applied strain can be easily computed as the area under the unloading curve. Thus, the energy dissipated is the difference between the area under the stress-strain curve minus the elastic energy. Thus, the fraction of energy dissipated has been plotted in Fig. \ref{fig:DissipatedE} as a function of the applied strain for the isotropic and anisotropic foams. The solid symbols stand for the experimental results while the open symbols come from the numerical simulations. Both experiments and simulations show that the fraction of energy dissipated increased with the applied strain and that it was always smaller in the isotropic foam as compared with the anisotropic foams, which presented the largest fraction of energy dissipated when deformed parallel to the longest axis of the cells. This conclusion emphasizes the large influence of the cell aspect ratio on the mechanical properties of the foams, that affects, up to large extent, not only the stiffness and the plateau stress but also the energy dissipated during deformation. It should be finally noted that the fraction of energy dissipated in the experiments was always higher than that calculated in the simulations. Energy was dissipated in the model through plastic deformation and friction but fracture of the cell walls and struts was not included and this may be the origin of the differences. Nevertheless, the differences between models and simulations were limited and the model was able to capture the experimental trends in terms of the effect of the anisotropy and strain on the fraction of energy dissipated.

\section{Analytical model} 

The simulation strategy presented in this paper has been used in  a previous investigation to develop surrogate models for the elastic modulus and the plateau stress of closed-cell foams taking into account the most relevant microstructural parameters \citep{marvi2018surrogate}. The information presented there can be used, in combination with the results for the evolution of the gas pressure within the cells during deformation in Fig. \ref{fig:NU}(b), to propose a simple analytical model to predict the mechanical response in compression of closed-cell foams taking into account the influence of the most relevant microstructural parameters. 

The analytical model assumes that the microstructure controls the elastic modulus and the plateau stress of the foam and that hardening during deformation is controlled by the increment in the gas pressure within the cells. This contribution can be approximated by eq. \eqref{eq:pres} assuming that the Poisson's ratio of the foam is 0. Thus,

\begin{equation}
\sigma = 
\begin{cases}  
E\epsilon  \quad & \mbox{if } \quad \epsilon \le \sigma_y/E  \\
\sigma_y + \Delta p(\epsilon) \quad & \mbox{if } \quad \epsilon \ge \sigma_y/E
\end{cases}
\label{Analytical}
\end{equation}

\noindent where the elastic modulus and the yield stress of the foam, $E$  and $\sigma_y$, respectively, are given by 
\citep{marvi2018surrogate},

\begin{equation}
\frac{E}{E_s} = C_1 s^{1.2} \bigg(\phi {\frac{\rho}{\rho_s}}\bigg)^{1.5}+C_{1}^\prime s^{1.1} (1-\phi)\bigg(\frac{\rho}{\rho_s}\bigg)^{1.2}
\label{eq:E}
\end{equation} 

\noindent and 

\begin{equation}
\frac{\sigma_{y}}{\sigma_{ys}} = C_3  s^{0.6} \phi^{3.5} \bigg({\frac{\rho}{\rho_s}}\bigg)^{1.5} + C_{4} s^{0.8} (1-\phi)\bigg(\frac{\rho}{\rho_s}\bigg)^{1.5}
\label{eq:S}
\end{equation}  

\noindent where $\phi$ stands for the fraction of material in the struts, $\rho_s$ the density of solid material in the foams, $s$ the cell aspect ratio and the coefficients $C_1$, $C_1^\prime$, $C_3$ and $C_4$  were obtained in \cite{marvi2018surrogate} by fitting the results of the numerical simulations of the foam behavior in compression for 0.025 $\le \rho/\rho_s \le 0.2$, $0.2 \le \phi < 1$ ($\phi$ = 1 stands for an open cell foam) and $0.5 \le s \le 2$. They are given in Table \ref{tab:c} for the sake of completion.

\begin{table}[h]
\begin{center}\label{tab:c}
\caption{Coefficients of eqs. \eqref{eq:E} and \eqref{eq:S}  to determine the elastic modulus and the plateau stress of closed-cell foams \cite{marvi2018surrogate}. }
\begin{tabular}{cccc}\\
\hline
  $C_1$ & $C_1^\prime$ &  $C_3$ & $C_4$  \\
 0.158 &   0.5155  & 0.156 & 0.760  \\
\hline
\end{tabular}
\end{center}
\end{table}

Finally, the pressure contribution, $\Delta p$, can be expressed as 

\begin{equation}
\Delta p = p_0\bigg[\frac{1-\rho/\rho_s}{(1-\rho/\rho_s) - \epsilon}-1\bigg].
\label{eq:pres0}
\end{equation} 

The predictions of eq. \eqref{Analytical} are compared with the experimental data and the simulation results for the isotropic and anisotropic foams in Figs. \ref{fig:IsoStressStrain} and \ref{fig:AnisoStressStrain}, respectively. They are in a very reasonable agreement with the experiments and simulations.Thus, this model provides a simple expression to make an  estimation of the mechanical properties of foams in compression including the most relevant microstructural parameters.

\begin{figure} [t!]
  \centering
  \includegraphics[scale=0.9]{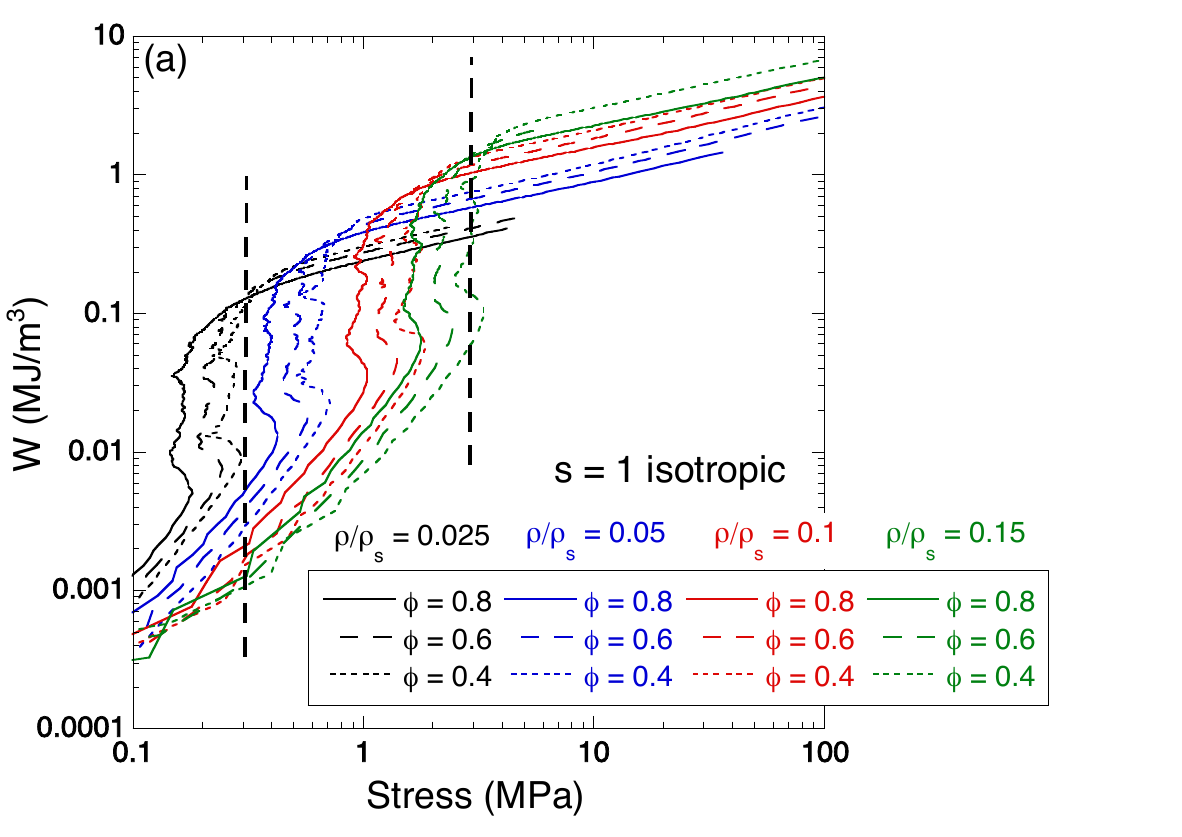}
   \includegraphics[scale=0.9]{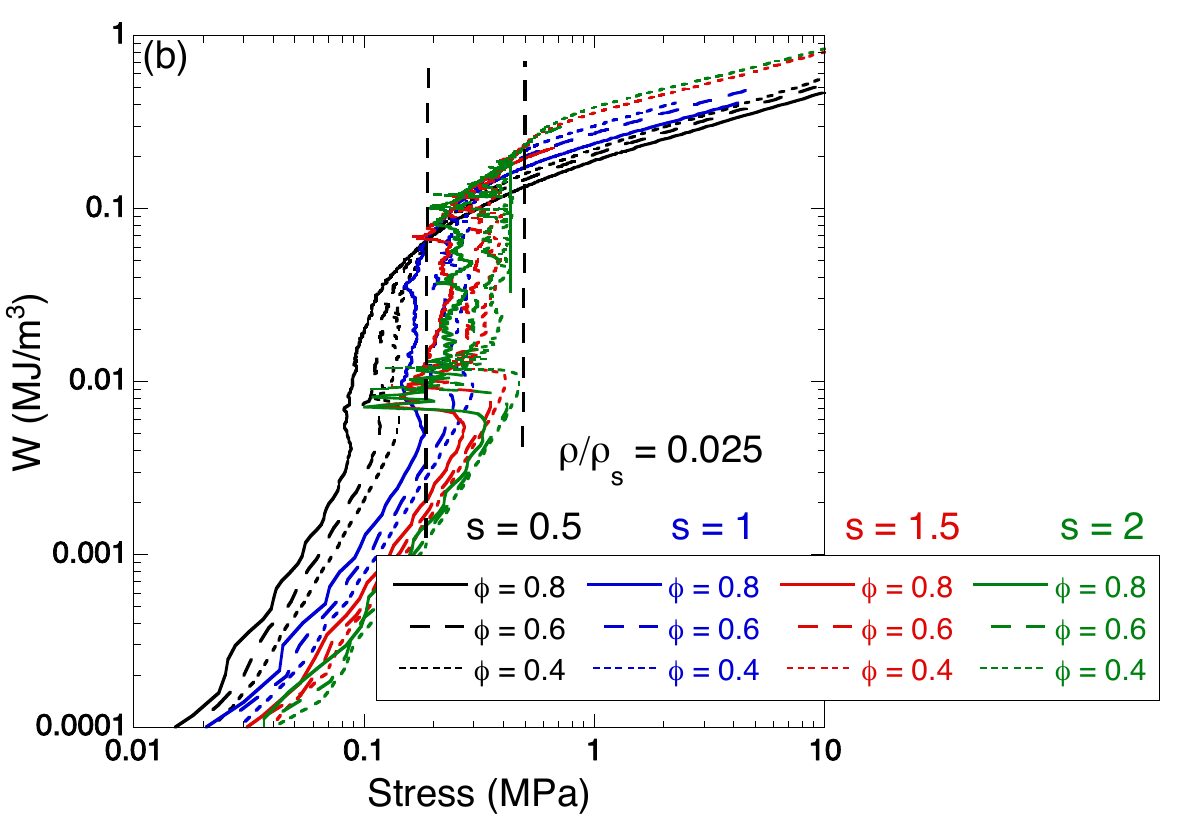}
  \caption{(a) Energy absorption {\it vs.} stress diagram for isotropic PU foams with different relative densities (0.025 $\le \rho/\rho_s \le $ 0.15) and fraction of mass in the struts ($0.4 \le \phi \le 0.8$). (b) Energy absorption {\it vs.} stress diagram for anisotropic PU foams with relative density  4 $\rho/\rho_s$ = 0.025, and different aspect ratios (0.5 $\le s <$ 2) and fraction of mass in the struts (0.4 $\le \phi \le$ 0.8).} 
  \label{fig:EneAbsDiagram}
\end{figure}

\section{Application for optimum design for packaging}

Foams used for packaging should absorb the maximum amount of energy during deformation but the stress carried by the foam should not be higher than a given value to avoid damage of the product. The simulation strategy presented above can be used to generate  diagrams that can be used to design the microstructure of foams with optimum properties for specific applications. This is the case of the energy absorbed - stress diagram plotted in Fig. \ref{fig:EneAbsDiagram}(a) for the case of isotropic foams. Each curve in the diagram stands for the energy absorbed (per unit volume) as a function of the stress carried by the foam and can be easily obtained from the compressive stress-strain curves obtained by the simulation of the RVE of the foam. The different curves in Fig. \ref{fig:EneAbsDiagram}(a) correspond to isotropic foams with relative densities in the range 0.025 $< \rho/\rho_s < $ 0.15 and a fraction of the mass in the struts in the range $0.4<\phi<0.8$. If the maximum stress carried by the foam cannot be higher than 0.3 MPa, as indicated by the corresponding dashed vertical line in Fig. \ref{fig:EneAbsDiagram}(a), the maximum energy absorbed is attained with the foams with the lowest density ($\rho/\rho_s$ = 0.025) and the maximum fraction of mass in the struts ($\phi$ = 0.8). Increasing the foam density in this regime (limited by the maximum stress carried by the foam) reduced the amount of energy absorbed because the limiting stress was attained before  reaching the plateau region in which the amount of energy absorbed increases rapidly with deformation at constant stress. On the contrary, if the maximum allowed stress is large (for instance, higher than 3 MPa as indicated by the corresponding dashed vertical line in the same figure), the best performance is attained with foams with the largest density ($\rho/\rho_s$ = 0.15) and the lowest fraction of mass in the struts ($\phi$ = 0.4). Energy absorption in this regime is controlled by the plateau stress, which increases with the density and the fraction of mass in the cell walls \citep{marvi2018surrogate}. When the maximum stress is in the range 0.3 - 3 MPa, the optimum foam is in between the two extreme cases and can be designed from the data in Fig. \ref{fig:EneAbsDiagram}(a).

Another important factor in the mechanical behavior of the foam is the cell anisotropy. The energy absorption diagram for  closed-cell foams with a relative density $\rho/\rho_s$ = 0.025 and different cell aspect ratios (0.5 $\le s \le$ 2) and fraction of the mass in the struts (0.4 $ \le \phi \le$ 0.8) is plotted in Fig. \ref{fig:EneAbsDiagram}(b). As in the previous case, two different regimes were found. In the low stress regime (maximum stress $<$ 0.2 MPa, see the corresponding dashed vertical line), the optimum performance from the viewpoint of energy absorption is obtained  with the foams with the lowest aspect ratio ($s$ = 0.5) and the highest fraction of mass in the struts ($\phi$ = 0.8). Both factors reduce the magnitude of the plateau stress \citep{marvi2018surrogate} and allowed the foam to absorb energy in the plateau region while the stress is below the limit. On the contrary, if the maximum stress is higher than 0.6 MPa (see the corresponding dashed line), the energy absorption increases with the foam aspect ratio and the fraction of mass in the cell walls. Once again, there is a transition between both regimes and the optimum foam in this case can be obtained from  the data in Fig. \ref{fig:EneAbsDiagram}(b).

\section{Conclusions}

The mechanical behavior in compression of closed-cell foams was simulated until full densification by means of the finite element analysis of a digital representative volume element of the microstructure of the foam. The digital representation includes all the relevant details of the foam microstructure (relative density, cell size distribution and shape, fraction of mass in the struts and cell walls, strut shape, etc.), while the numerical simulation takes into account the influence of the gas pressure in the cells and of the contact between cell walls and struts during crushing.

The model was validated by comparison with experimental results on isotropic and anisotropic PU foams and it was able to reproduce accurately the initial stiffness, the plateau stress and the hardening region until full densification. Moreover, it also showed that the effect of friction was negligible during crushing and provided good estimations of the energy dissipated and of the elastic energy stored in the foam as  a function of the applied strain. Based on the simulation results, a simple analytical model was developed to predict the mechanical behavior of closed-cell foam in compression until very large strains ($<$ 0.8) before densification becomes dominant. The model includes the effect of the microstructure through the elastic modulus and the plateau stress that are obtained from accurate surrogate models developed previously using the same strategy \citep{marvi2018surrogate} and accounts for the influence of the gas pressure during densification by assuming that the PoissonÕs ratio of the foam during deformation is 0. Finally, an example of application of the simulation tool is presented to design foams with an optimum microstructure from the viewpoint of energy absorption for packaging.

\section*{Acknowledgements}
The development of the simulation strategy presented in this paper was supported  by the 7th Framework Programme of the European union within the framework the MODENA project (MOdelling of morphology DEvelopment of micro- and NAnostructures), contract number 604271.  

\section*{Supplementary information}

The simulation strategy presented in this paper is implemented in the software tool MULTIFOAM. More information about MULTIFOAM can be obtained at  https://materials.imdea.org/multifoam/

%\bibliographystyle{elsarticle-harv}
%\bibliography{mybibfile}

\begin{thebibliography}{30}
\expandafter\ifx\csname natexlab\endcsname\relax\def\natexlab#1{#1}\fi
\expandafter\ifx\csname url\endcsname\relax
  \def\url#1{\texttt{#1}}\fi
\expandafter\ifx\csname urlprefix\endcsname\relax\def\urlprefix{URL }\fi

\bibitem[{Abaqus(2016)}]{A16}
Abaqus, 2016. Analysis User's Manual, version 6.13. Dassault Systemes.

\bibitem[{Avalle et~al.(2007)Avalle, Belingardi, and Ibba}]{ABI07}
Avalle, M., Belingardi, G., Ibba, A., 2007. Mechanical models of cellular
  solids: {P}arameter identification from experimental tests. International
  Journal of Impact Engineering 34, 3 -- 27.

\bibitem[{Avalle et~al.(2001)Avalle, Belingardi, and Montanini}]{Avalle2001455}
Avalle, M., Belingardi, G., Montanini, R., 2001. Characterization of polymeric
  structural foams under compressive impact loading by means of
  energy-absorption diagram. International Journal of Impact Engineering, 25,
  455 -- 472.

\bibitem[{Berlin(1980)}]{Berlin1980}
Berlin, A., 1980. Chemistry and Processing of Foamed Polymers. Springer Berlin
  Heidelberg.

\bibitem[{Chen et~al.(2015)Chen, Das, and Battley}]{Chen2015150}
Chen, Y., Das, R., Battley, M., 2015. Effects of cell size and cell wall
  thickness variations on the stiffness of closed-cell foams. International
  Journal of Solids and Structures 52, 150 -- 164.

\bibitem[{Daniel et~al.(2009)Daniel, Gdoutos, and Rajapakse}]{Daniel2009}
Daniel, I.~M., Gdoutos, E.~E., Rajapakse, Y.~D., 2009. Major accomplishments in
  composite materials and sandwich structures: an anthology of ONR sponsored
  research. Springer Science \& Business Media.

\bibitem[{Geuzaine and Remacle(2009)}]{GMsh}
Geuzaine, C., Remacle, J.-F., 2009. Gmsh: A {3-D} finite element mesh generator
  with built-in pre- and post-processing facilities. International Journal for
  Numerical Methods in Engineering 79, 1309--1331.

\bibitem[{Gibson and Ashby(1999)}]{GibsonBookCellular}
Gibson, L.~J., Ashby, 1999. Cellular Solids: Structure and Properties.
  Cambridge University Press.

\bibitem[{Jang et~al.(2008)Jang, Kraynik, and Kyriakides}]{Jang20081845}
Jang, W.-Y., Kraynik, A.~M., Kyriakides, S., 2008. On the microstructure of
  open-cell foams and its effect on elastic properties. International Journal
  of Solids and Structures 45, 1845 -- 1875.

\bibitem[{Jang et~al.(2010)Jang, Kyriakides, and Kraynik}]{Jang20102872}
Jang, W.-Y., Kyriakides, S., Kraynik, A.~M., 2010. On the compressive strength
  of open-cell metal foams with kelvin and random cell structures.
  International Journal of Solids and Structures 47, 2872 -- 2883.

\bibitem[{Kanit et~al.(2003)Kanit, Forest, Galliet, Mounoury, and
  Jeulin}]{kanit2003determination}
Kanit, T., Forest, S., Galliet, I., Mounoury, V., Jeulin, D., 2003.
  Determination of the size of the representative volume element for random
  composites: statistical and numerical approach. International Journal of
  Solids and Structures 40, 3647--3679.

\bibitem[{K\"oll and Hallstr\"om(2016)}]{KH16}
K\"oll, J., Hallstr\"om, S., 2016. Elastic properties of equilibrium foams.
  Acta Materialia 113, 11 -- 18.

\bibitem[{Lautensack(2007)}]{L2007}
Lautensack, C., 2007. Random laguerre tessellations. {PhD} thesis,
  Universit\"at Karlsruhe.

\bibitem[{Liu et~al.(2005)Liu, Subhash, and Gao}]{LSG05}
Liu, Q., Subhash, G., Gao, X.~L., 2005. Parametric study on crushability of
  open-cell structural polymeric foams. Journal of Porous Materials 12, 233 --
  248.

\bibitem[{Marvi-Mashhadi et~al.(2018{\natexlab{a}})Marvi-Mashhadi, Lopes, and
  LLorca}]{MLL17anisotropy}
Marvi-Mashhadi, M., Lopes, C., LLorca, J., 2018{\natexlab{a}}. Effect of
  anisotropy on the mechanical properties of polyurethane foams: An
  experimental and numerical study. Mechanics of Materials 124, 143--154.

\bibitem[{Marvi-Mashhadi et~al.(2018{\natexlab{b}})Marvi-Mashhadi, Lopes, and
  LLorca}]{MLL17}
Marvi-Mashhadi, M., Lopes, C., LLorca, J., 2018{\natexlab{b}}. Modelling of the
  mechanical behavior of polyurethane foams by means of micromechanical
  characterization and computational homogenization. International Journal of
  Solids and Structures 146, 154 -- 166.

\bibitem[{Marvi-Mashhadi et~al.(2018{\natexlab{c}})Marvi-Mashhadi, Lopes, and
  LLorca}]{marvi2018surrogate}
Marvi-Mashhadi, M., Lopes, C.~S., LLorca, J., 2018{\natexlab{c}}. Surrogate
  models of the influence of the microstructure on the mechanical properties of
  closed-and open-cell foams. Journal of Materials Science 53, 12937--12948.

\bibitem[{Mills(2007)}]{Mills2007xvii}
Mills, N., 2007. Polymer Foams Handbook. Butterworth-Heinemann, Oxford.

\bibitem[{Mills et~al.(2009)Mills, Stampfli, Marone, and Br\"uhwiler}]{MSM09}
Mills, N.~J., Stampfli, R., Marone, F., Br\"uhwiler, P.~A., 2009. Finite
  element micromechanics model of impact compression of closed-cell polymer
  foams. International Journal of Solids and Structures 46, 677 -- 697.

\bibitem[{Ozturk and Anlas(2009)}]{OA09}
Ozturk, U.~E., Anlas, G., 2009. Energy absorption calculations in multiple
  compressive loading of polymeric foams. Materials \& Design 30, 15 -- 22.

\bibitem[{Quey et~al.(2011)Quey, Dawson, and Barbe}]{NEPER}
Quey, R., Dawson, P., Barbe, F., 2011. Large-scale 3d random polycrystals for
  the finite element method: Generation, meshing and remeshing. Computer
  Methods in Applied Mechanics and Engineering 200, 1729 -- 1745.

\bibitem[{Redenbach et~al.(2012)Redenbach, Shklyar, and
  Andr{\"a}}]{Redenbach201270}
Redenbach, C., Shklyar, I., Andr{\"a}, H., 2012. Laguerre tessellations for
  elastic stiffness simulations of closed foams with strongly varying cell
  sizes. International Journal of Engineering Science 50, 70 -- 78.

\bibitem[{Rodriguez-Perez et~al.(2009)Rodriguez-Perez, Hidalgo, Solorzano, and
  de~Saja}]{rodriguez2009measuring}
Rodriguez-Perez, M., Hidalgo, F., Solorzano, E., de~Saja, J., 2009. Measuring
  the time evolution of the gas pressure in closed cell polyolefin foams
  produced by compression moulding. Polymer Testing 28, 188--195.

\bibitem[{S{\'a}daba et~al.(2019)S{\'a}daba, Herr{\'a}ez, Naya, Gonz{\'a}lez,
  Llorca, and Lopes}]{sadaba2019special}
S{\'a}daba, S., Herr{\'a}ez, M., Naya, F., Gonz{\'a}lez, C., Llorca, J., Lopes,
  C., 2019. Special-purpose elements to impose periodic boundary conditions for
  multiscale computational homogenization of composite materials with the
  explicit finite element method. Composite Structures 208, 434--441.

\bibitem[{Segurado and LLorca(2002)}]{SL02}
Segurado, J., LLorca, J., 2002. A numerical approximation to the elastic
  properties of sphere-reinforced composites. Journal of the Mechanics and
  Physics of Solids 50, 2107--2121.

\bibitem[{Subramanian and Sankar(2012)}]{Subramanian25052012}
Subramanian, N., Sankar, B.~V., 2012. Evaluation of micromechanical methods to
  determine stiffness and strength properties of foams. Journal of Sandwich
  Structures \& Materials 14~(4), 431--447.

\bibitem[{Sullivan et~al.(2008)Sullivan, Ghosn, and Lerch}]{Sullivan20081754}
Sullivan, R.~M., Ghosn, L.~J., Lerch, B.~A., 2008. A general tetrakaidecahedron
  model for open-celled foams. International Journal of Solids and Structures
  45, 1754 -- 1765.

\bibitem[{Tekog˜lu et~al.(2011)Tekog˜lu, Gibson, Pardoen, and
  Onck}]{TEKOGLU2011109}
Tekog˜lu, C., Gibson, L.~J., Pardoen, T., Onck, P.~R., 2011. Size effects in
  foams: Experiments and modeling. Progress in Materials Science 56, 109 --
  138.

\bibitem[{Vecchio et~al.(2016)Vecchio, Redenbach, Schladitz, and
  Kraynik}]{VRS16}
Vecchio, I., Redenbach, C., Schladitz, K., Kraynik, A., 2016. Improved models
  of solid foams based on soap froth. Computational Materials Science 120,
  60--69.

\bibitem[{Youssef et~al.(2005)Youssef, Maire, and Gaertner}]{YMG05}
Youssef, S., Maire, E., Gaertner, R., 2005. Finite element modelling of the
  actual structure of cellular materials determined by x-ray tomography. Acta
  Materialia 53, 719 -- 730.

\end{thebibliography}

\end{document}